\documentclass[aps,preprint,superscriptaddress,showpacs,amsmath,amssymb,endfloats*]{revtex4}
\usepackage{graphicx}
\usepackage{color}
\usepackage{multibib}
\usepackage[normalem]{ulem}
\begin{document}

\title{Magnetically driven superconductivity in \boldmath{CeCu$_2$Si$_2$}}

\author{O.~Stockert} 
\email{stockert@cpfs.mpg.de}
\affiliation{Max-Planck-Institut f\"ur Chemische Physik fester Stoffe, 01187 Dresden, Germany} 
\author{J.~Arndt}
\affiliation{Max-Planck-Institut f\"ur Chemische Physik fester Stoffe, 01187 Dresden, Germany} 
\author{E.~Faulhaber}
\affiliation{Institut f\"ur Festk\"orperphysik, Technische Universit\"at
  Dresden, 01062 Dresden, Germany}
\affiliation{Gemeinsame Forschergruppe Helmholtz-Zentrum Berlin - TU Dresden, 85747 Garching, Germany}
\author{C.~Geibel} 
\affiliation{Max-Planck-Institut f\"ur Chemische Physik fester Stoffe, 01187 Dresden, Germany} 
\author{H.~S.~Jeevan} 
\affiliation{Max-Planck-Institut f\"ur Chemische Physik fester Stoffe, 01187 Dresden, Germany} 
\affiliation{I. Physikalisches Institut, Universit\"at G\"ottingen, 37077 G\"ottingen, Germany}
\author{S.~Kirchner}
\affiliation{Max-Planck-Institut f\"ur Chemische Physik fester Stoffe, 01187 Dresden, Germany} 
\affiliation{Max-Planck-Institut f\"ur Physik komplexer Systeme, 01187 Dresden, Germany}
\author{M.~Loewenhaupt}
\affiliation{Institut f\"ur Festk\"orperphysik, Technische Universit\"at
  Dresden, 01062 Dresden, Germany}
\author{K.~Schmalzl}
\affiliation{J\"ulich Centre for Neutron Science at Institut Laue-Langevin, 38042 Grenoble, France}
\author{W.~Schmidt}
\affiliation{J\"ulich Centre for Neutron Science at Institut Laue-Langevin, 38042 Grenoble, France}
\author{Q.~Si}
\affiliation{Department of Physics and Astronomy, Rice University, Houston, Texas, 77005, USA}
\author{F.~Steglich} 
\affiliation{Max-Planck-Institut f\"ur Chemische Physik fester Stoffe, 01187 Dresden, Germany}

\date{\today}

\begin{abstract}
\end{abstract}


\maketitle

{\bf \boldmath
The origin of unconventional superconductivity,
including high-temperature and heavy-fermion superconductivity,
is still a matter of controversy.
Spin excitations instead of phonons are thought to be responsible for 
the formation of Cooper pairs.
Using inelastic neutron scattering, we present the first in-depth study of the magnetic excitation spectrum in momentum and energy space in the superconducting and the normal states of 
CeCu$_2$Si$_2$. A clear spin excitation gap is observed in the superconducting state.
We determine a lowering of the magnetic exchange energy in the superconducting state, 
in an amount considerably larger than the superconducting 
condensation energy. Our findings identify the antiferromagnetic 
excitations as the major driving force for superconducting pairing in
this prototypical heavy-fermion compound 
located near an antiferromagnetic quantum critical point.
}

While conventional superconductivity (SC) is generally incompatible 
with magnetism, magnetic excitations seem to play an important role 
in the Cooper pair formation of unconventional superconductors such as the
high-$T_c$ cuprates or the low-$T_c$ organic and heavy-fermion (HF)
superconductors. Since the discovery of SC in 
CeCu$_2$Si$_2$ \cite{Steglich.79}, antiferromagnetic (AF) spin excitations 
have been proposed as a viable mechanism for 
SC \cite{Miyake.86,Scalapino.86,Monthoux.07}. 
The discovery of SC at the 
boundary of AF order in CePd$_2$Si$_2$ \cite{Mathur.98}
has pushed this notion into the framework of AF quantum criticality~\cite{Gegenwart.08}.
Unfortunately, such quantum critical points (QCPs) proximate to HF
superconductors typically arise under pressure, which 
makes it difficult to probe their magnetic excitation spectrum. 

\begin{figure}[b]
\includegraphics[clip,width=\linewidth]{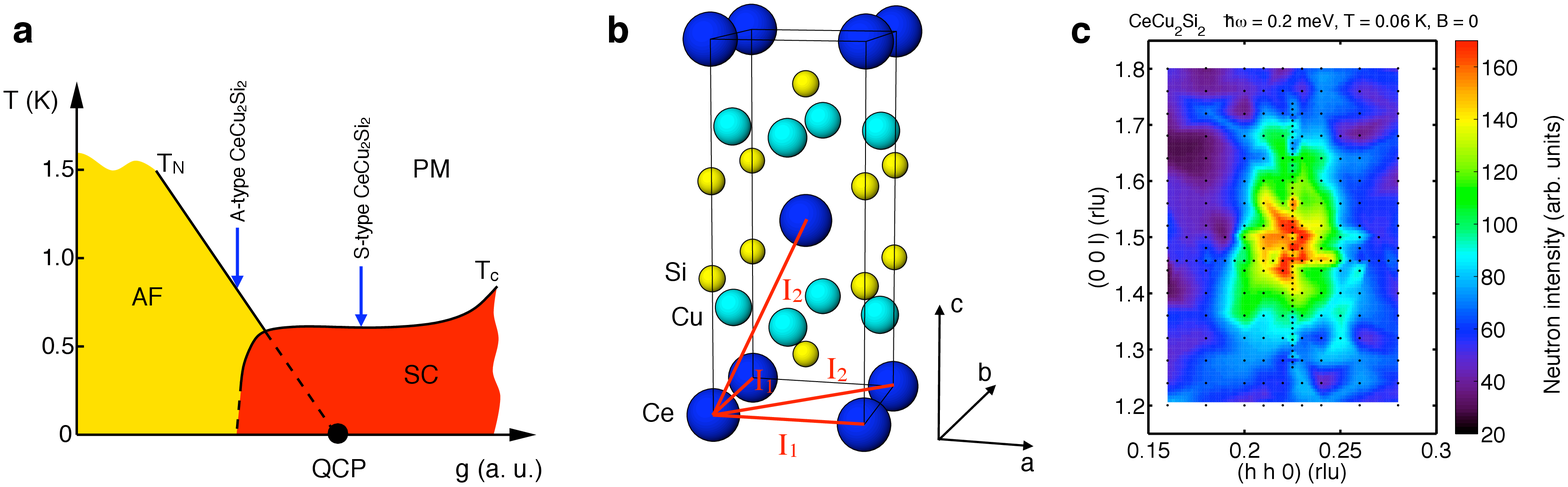}
\caption{{\bf Schematic phase diagram around the QCP, crystal structure and nearly isotropic spin fluctuations of CeCu$_2$Si$_2$.} (a) Schematic $T-g$ phase diagram of CeCu$_2$Si$_2$ in the vicinity 
of the quantum critical point (QCP) where the antiferromagnetic (AF) 
phase vanishes as function of the effective coupling constant $g$. 
Superconductivity (SC) is observed around the QCP and extends far 
into the paramagnetic (PM) regime. Composition as well as hydrostatic 
pressure can be used to change the coupling constant $g$ and to tune 
the system to the QCP. The positions of the A-type and the S-type 
single crystals in the phase diagram are marked. (b) Tetragonal crystal structure (space group: $I4/mmm$) of CeCu$_2$Si$_2$. The nearest and next-nearest neighbour interactions between the cerium atoms are labelled by $I_1$ and $I_2$. It should be noted that the distances between next-nearest neighbour Ce atoms in the basal plane and out-of-plane are almost identical.
(c) The spin fluctuations at $T = 0.06$\,K and $B=0$ and at an energy transfer
$\hbar\omega = 0.2$\,meV.
The anisotropy factor between  the $[110]$ and the $[001]$ directions is about 1.5.
Note that the correct aspect ratio $[110]^{*}:[001]^{*}$ has been taken into account although the axes are labelled in reciprocal
lattice units (rlu). Black dots mark the $({\bf Q},\omega)$ positions data were taken.}
\label{phasediagram}
\end{figure}

Here, we report a detailed study of the magnetic excitations in
CeCu$_2$Si$_2$, which exhibits SC below $T_c \approx 0.6$\,K.
This prototypical HF compound is ideally suited for our purpose,
since SC here is in proximity to  
an AF QCP already at ambient pressure (cf. Fig.~\ref{phasediagram}(a)). 
As displayed in Fig.~\ref{phasediagram}(b) CeCu$_2$Si$_2$ crystallises in a structure with body-centred tetragonal symmetry and is one of the best studied HF superconductors and well characterised by low-temperature transport and thermodynamic measurements \cite{Steglich.01}.
Moreover, those measurements in the 
field-induced normal state have already provided evidence 
that the QCP in this compound is of the
three-dimensional (3D) spin-density-wave (SDW) type~\cite{Gegenwart.98}.
The spatial anisotropy of the spin fluctuations in superconducting
CeCu$_2$Si$_2$ was measured at $T = 0.06$\,K and at an energy transfer
$\hbar\omega = 0.2$\,meV and is shown in Fig.~\ref{phasediagram}(c). These magnetic  
correlations
display only a small anisotropy (a factor of $1.5$) in the correlation
lengths between the $[110]$ and the $[001]$ direction. Therefore,
these quite isotropic spin fluctuations are in line with  
thermodynamic and
transport measurements exhibiting $C/T = \gamma_0 - a \sqrt{T}$ or
$\rho - \rho_0 = A T^\alpha$, $\alpha = 1-1.5$ \cite{Yuan.03,Gegenwart.98},
and strongly support a three-dimensional quantum critical SDW scenario \cite{Rosch.99}.
We are able to identify the magnetic excitations in the normal
state of paramagnetic, superconducting CeCu$_2$Si$_2$, 
around the incommensurate wave vector \cite{Stockert.04} 
of the SDW order nearby in the phase diagram (cf. Fig.~\ref{phasediagram}(a)), 
and further establish the system's proximity to the AF QCP through
the observation of a considerable slowing down in the spin dynamics. Going into the superconducting 
state, 
a spin gap opens out of a broadened quasielastic response which extends
to much higher frequencies 
(10 $\times$ the superconducting gap).
These data allow us to establish a saving in the AF exchange energy
that is considerably larger than the superconducting condensation
energy, thereby providing the first demonstration 
of the nearly-quantum-critical AF excitations
as the major driving force for unconventional SC.

\subsection*{Superconductivity and antiferromagnetism in CeCu$_2$Si$_2$}
The SC in CeCu$_2$Si$_2$ we consider is close to the
AF QCP, and is to be contrasted with a second superconducting dome
appearing at high pressure which is thought to be associated with 
a valence instability and the concomitant 
fluctuations~\cite{Yuan.03,Holmes.04}.
This AF QCP  is located within the narrow homogeneity range of
the "122" phase in the ternary chemical Ce-Cu-Si phase diagram
of this tetragonal compound \cite{Steglich.96}. Correspondingly,
we can prepare {\it homogeneous} 
samples (with slight Cu deficit) from the antiferromagnetically ordered 
side (A-type) and (with tiny Cu excess) from the paramagnetic, 
superconducting  side of the QCP (S-type);
by contrast, crystals very close 
to the $1:2:2$ stoichiometry exhibit a ground state where SC and AF 
compete with each other without microscopic coexistence (A/S-type) \cite{Steglich.96}.
The AF order was found to be an incommensurate SDW \cite{Stockert.04}.
 At $T = 0.05$\,K, well below $T_{\rm N} \approx 0.8$\,K,
the A-type CeCu$_2$Si$_2$ exhibits an ordered magnetic moment 
$\mu_{ord} \approx 0.1\,\mu_{\rm B}$ and an incommensurate propagation 
vector $\tau \approx (0.215~0.215~0.53)$. 
The latter can be ascribed to a nesting wave vector of the renormalised Fermi surface.
However, a full microscopic description of the magnetic order remains to be addressed.

In order to study the superconducting state in CeCu$_2$Si$_2$ in detail, neutron 
scattering results \cite{natureonline} on an S-type single crystal are presented 
in this Article. A previous experiment was severely hampered by a large experimental
 background and a low signal-to-background ratio \cite{Stockert.08}. Thermodynamic 
measurements confirmed that this crystal is superconducting with a 
$T_c \approx 0.6$\,K and an upper critical field $B_{c2}(T = 0) < 2$\,T \cite{natureonline}. 
Elastic neutron scattering measurements did not feature resolution-limited 
magnetic Bragg peaks in S-type CeCu$_2$Si$_2$ in accordance with thermodynamic 
measurements. However, at positions where magnetic satellite peaks are observed
 in A-type CeCu$_2$Si$_2$ \cite{Stockert.04}, e.g., at ${\bf Q_{\rm AF}} = (0.215~0.215~1.458)$, 
relative to a nearby nuclear Bragg reflection ${\bf G}$ (${\bf Q_{\rm AF}} = {\bf G}\pm \tau$),
the S-type crystal exhibits quite weak correlation peaks at low temperatures 
\cite{Stockert.08}. They are still present above $T_c$ and disappear at 
$T \approx 0.8$\,K, very similar to the behaviour of the SDW order in 
A-type CeCu$_2$Si$_2$ \cite{Stockert.04}.
Although these peaks were found to be purely elastic within the energy 
resolution ($\approx 57$\,$\mu$eV), their linewidth in ${\bf Q}$ space is considerably broadened 
corresponding to a correlation length of $50-60$\,{\AA} being comparable to the 
superconducting coherence length of order $100$\,{\AA} \cite{Rauchschwalbe.82}. 
Thus, static magnetically ordered regions seem to exist in a quite small part of the sample and are separated from the surrounding superconducting regions.

\subsection*{Spin dynamics in CeCu$_2$Si$_2$}
\begin{figure*}[t]
\includegraphics[clip,width=0.55\linewidth]{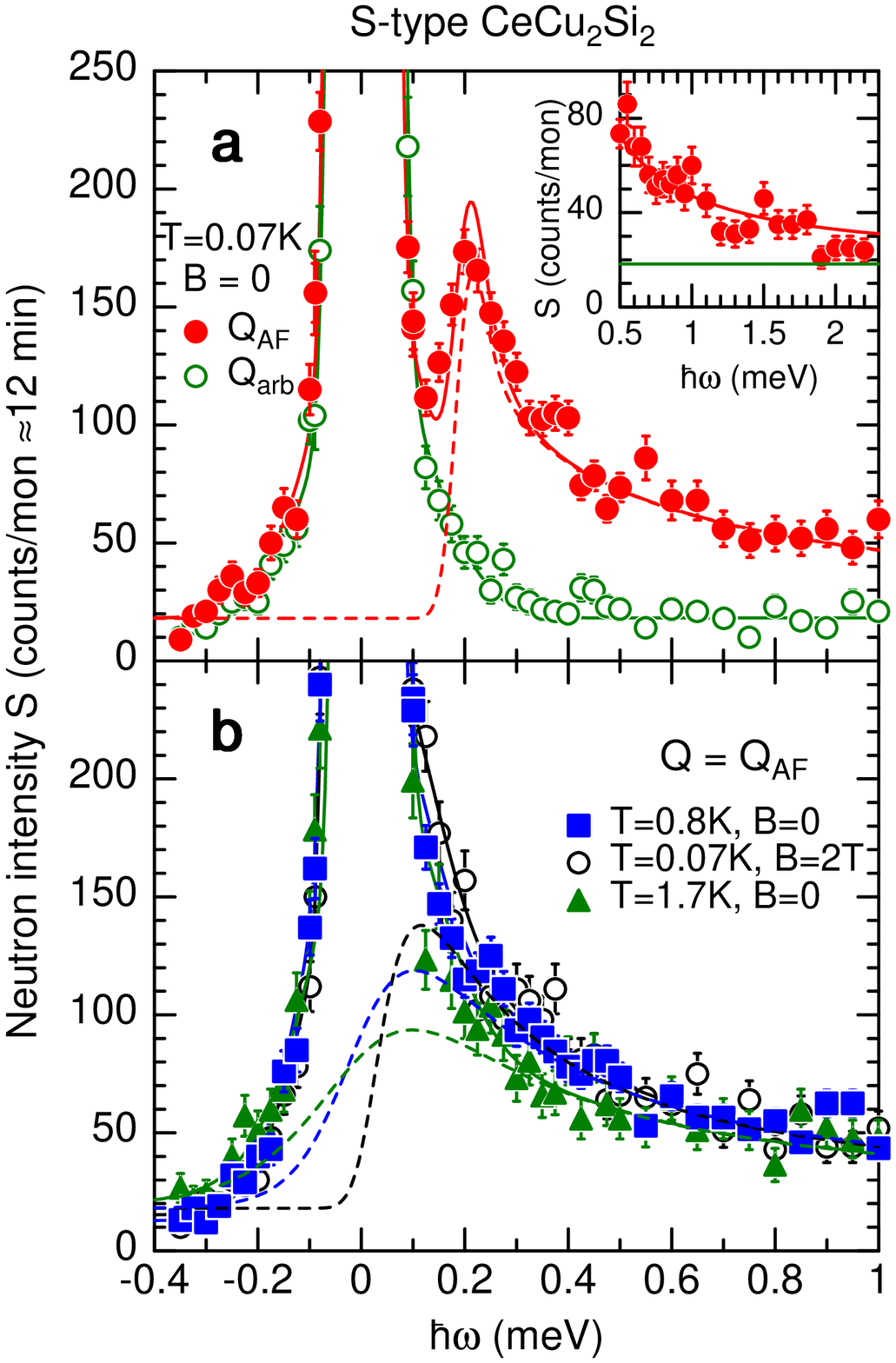}
\caption{{\bf Inelastic neutron scattering spectra in the normal and superconducting states of CeCu$_2$Si$_2$}. Energy scans (neutron intensity $S = S_{ela} + S_{qe/ine, mag}$ versus energy 
transfer $\hbar\omega$) in S-type CeCu$_2$Si$_2$ at 
${\bf Q} = {\bf Q_{\rm AF}} = (0.215~0.215~1.458)$ in (a) the superconducting 
state at $T = 0.07$\,K, $B = 0$ and in (b) the normal state at $T = 0.8$ and 
$1.7$\,K, $B = 0$ and $T = 0.07$\,K, $B = 2$\,T. For comparison the magnetic 
response at an arbitrary, general ${\bf Q}$ position 
${\bf Q} = {\bf Q_{\rm arb}} = (0.1~0.1~1.6)$ at $T = 0.07$\,K, $B = 0$ 
is also plotted in (a). The inset in (a) shows the magnetic response at 
$\bf Q_{\rm AF}$ ($T = 0.07$\,K, $B = 0$) extending beyond $\hbar\omega = 2$\,meV.
Solid lines represent fits to the data comprised of the incoherent and coherent 
elastic signal $S_{ela}$ and the quasielastic/gapped inelastic magnetic 
response $S_{qe/ine, mag}$ (dashed lines) with Lorentzian lineshape convolved with the 
resolution. The gapped magnetic response at $T = 0.07$\,K, $B = 0$ has 
been modelled by a quasielastic Lorentzian line taking into account a 
spin gap with a value $\hbar\omega_{\rm gap}$ and an enhanced density 
of states above the gap as for the electronic gap of a BCS superconductor.
The error bars represent the statistical error.}
\end{figure*}
\begin{figure*}[t]
\includegraphics[clip,width=0.55\linewidth]{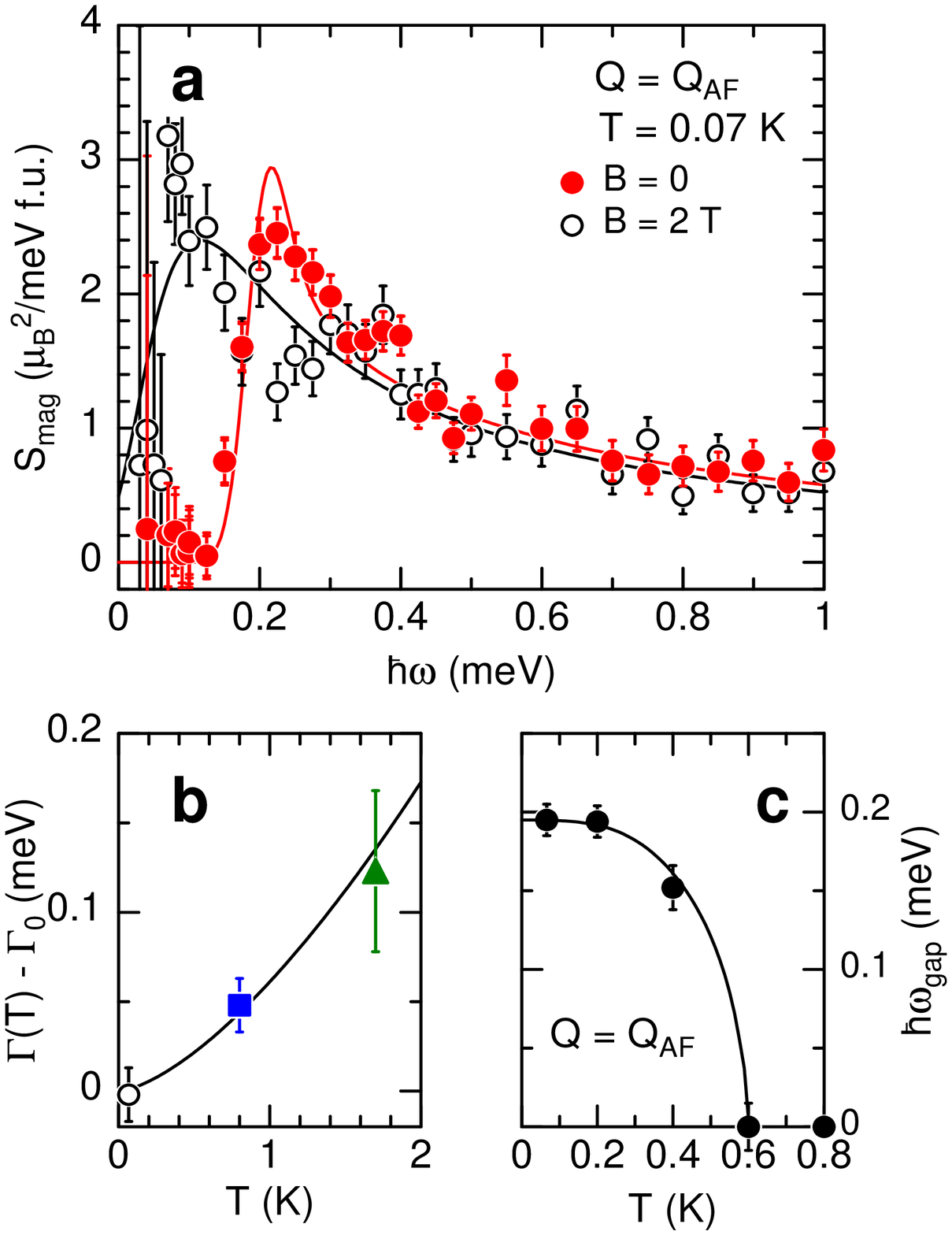}
\caption{{\bf Magnetic response, relaxation rate and spin gap at the AF wave vector of CeCu$_2$Si$_2$.}
(a) Magnetic response $S_{qe/ine, mag}$ at ${\bf Q_{\rm AF}}$ and 
$T = 0.07$\,K in the superconducting and the normal state, i.e., at $B = 0$ 
and $B = 2$\,T,  as extracted from the data displayed in Fig.\,2. The data 
have been put on an absolute intensity scale \cite{natureonline}. 
Below $\hbar\omega \approx 0.1$\,meV ($\approx 1.6 \times$ the instrumental resolution) the errors in $S_{mag}$ increase strongly
(some data points fall even outside the plotted range), 
since the strong elastic scattering $S_{ela}$ is subtracted from the total scattering to receive $S_{mag}$ and due to small uncertainties in the resolution function. These uncertainties are the same for both data sets and do not play a role since only the difference is analysed for the estimation of the exchange energy saving.
(b) Linewidth $\Gamma$ vs. temperature $T$ of 
the quasielastic magnetic response at ${\bf Q_{\rm AF}}$ in the normal state as
yielded by fits to the data shown in Fig.\,2.
Plotted here is $\Gamma(T)-\Gamma_0$ vs. $T$,
with $\Gamma_0=0.112\mbox{meV}$.
The solid line $\Gamma(T)-\Gamma_0=a\,T^{3/2}$ (with $a=0.061$\,meV/K$^{1.5}$) 
is the expected behaviour near a 3D SDW QCP.
(c) Temperature dependence of the spin excitation gap $\hbar\omega_{\rm gap}$ 
at ${\bf Q_{\rm AF}}$ together with the scaled d-wave BCS superconducting 
gap function (solid line). The error bars denote the statistical error.}
\end{figure*}

We probe the magnetic response of CeCu$_2$Si$_2$ through extensive inelastic neutron scattering measurements around 
${\bf Q} = {\bf Q_{\rm AF}} = (0.215~0.215~1.458)$, since no appreciable magnetic intensity has been detected elsewhere in the Brillouin zone.
Fig. 2(a) displays energy scans at this ${\bf Q_{\rm AF}}$ position and at a 
general position ${\bf Q} = {\bf Q_{\rm arb}} = (0.1~0.1~1.6)$, where no correlation
 peaks emerge, but which has the same $|{\bf Q}|$. Both data sets were recorded
 in the superconducting state at $T = 0.07$\,K. At ${\bf Q_{\rm arb}}$ only the
 incoherent elastic background contribution with instrument resolution is seen,
 while no magnetic intensity could be detected. In contrast, at ${\bf Q_{\rm AF}}$
 the response shows a strong inelastic signal with a long tail of the 
intensity extending beyond $\hbar\omega = 2$\,meV (cf. inset of Fig.\,2(a)).
The missing spectral weight at low energies is an indication for a 
spin-excitation gap in the superconducting state. 
The spectrum recovers the missing weight 
at the gap edge, thereby constituting an inelastic line.
The data can be described by a quasielastic Lorentzian line with a spin 
excitation gap $\hbar\omega_{gap} \approx 0.2$\,meV and with a density of
 states as for the electronic gap of a d-wave BCS  superconductor 
(solid lines in Figs. 2(a) and 3(a)).
$\hbar\omega_{\mbox{gap}} \approx 3.9 k_B T_c$ is found to be $10 \%$ smaller than the value 
predicted for a weak-coupling d-wave superconductor \cite{Ohkawa.87} and falls $20\%$ below
$2\Delta_0/k_B T_c=5.0$ as determined by Cu-NQR for CeCu$_2$Si$_2$~\cite{Ishida.99,Fujiwara.08}.
To unambiguously relate the inelastic magnetic excitation to the 
superconducting state,
it was necessary to perform additional measurements in the normal state.

Energy scans recorded at $\bf Q_{\rm AF}$ in the normal state are shown in Fig. 2(b).
Notably, independent of how the normal state is reached, i.e., above $T_c$ 
at $T = 0.8$\,K and $B = 0$ or above $B_{c2}$ at $T = 0.07$\,K and $B = 2$\,T, 
the magnetic response is almost identical and appears to be quasielastic. 
The fits to the quasielastic magnetic response  with a Lorentzian lineshape
 give a good description of the data as seen in Figs. 2(b) and 3(a). 
With increasing temperature in the normal state the magnetic response 
weakens in intensity and broadens considerably. Starting from 
$\Gamma \approx 0.11$\,meV at $T = 0.07$\,K the linewidth of the quasielastic
 response at ${\bf Q_{\rm AF}}$ increases to $\Gamma \approx 0.235$\,meV at
 $T = 1.7$\,K (Fig. 3(b)). This considerable slowing down of the response 
when lowering the temperature indicates the proximity of S-type CeCu$_2$Si$_2$
 to the AF QCP. 
$\Gamma(T)$ extrapolates to a finite value at $T\rightarrow 0$,
since the S-type single crystal is located on
the paramagnetic side of the QCP (cf. Fig. 1(a)).
A related critical slowing down was observed 
in magnetically ordered A-type CeCu$_2$Si$_2$ \cite{Stockert.06a}.

The fact that the magnetic excitation gap disappears in the normal state, i.e.,
 above $T_c$, 
and also above $B_{c2}$ at low temperatures, where the magnetic short-range 
correlations still persist,
gives direct evidence that the 
spin gap $\hbar\omega_{\rm gap}$ is related to the superconducting state. 
Its temperature variation is displayed in Fig. 3(c) and has been derived from
fits to the data shown in Figs. 2(a) and 3(a) and additional scans. As indicated
by the solid line, $\hbar\omega_{\rm gap}$ follows, within the error bars, 
the BCS form for the superconducting gap amplitude $2\Delta(T)$.

\begin{figure}[t]
\includegraphics[clip,width=\linewidth]{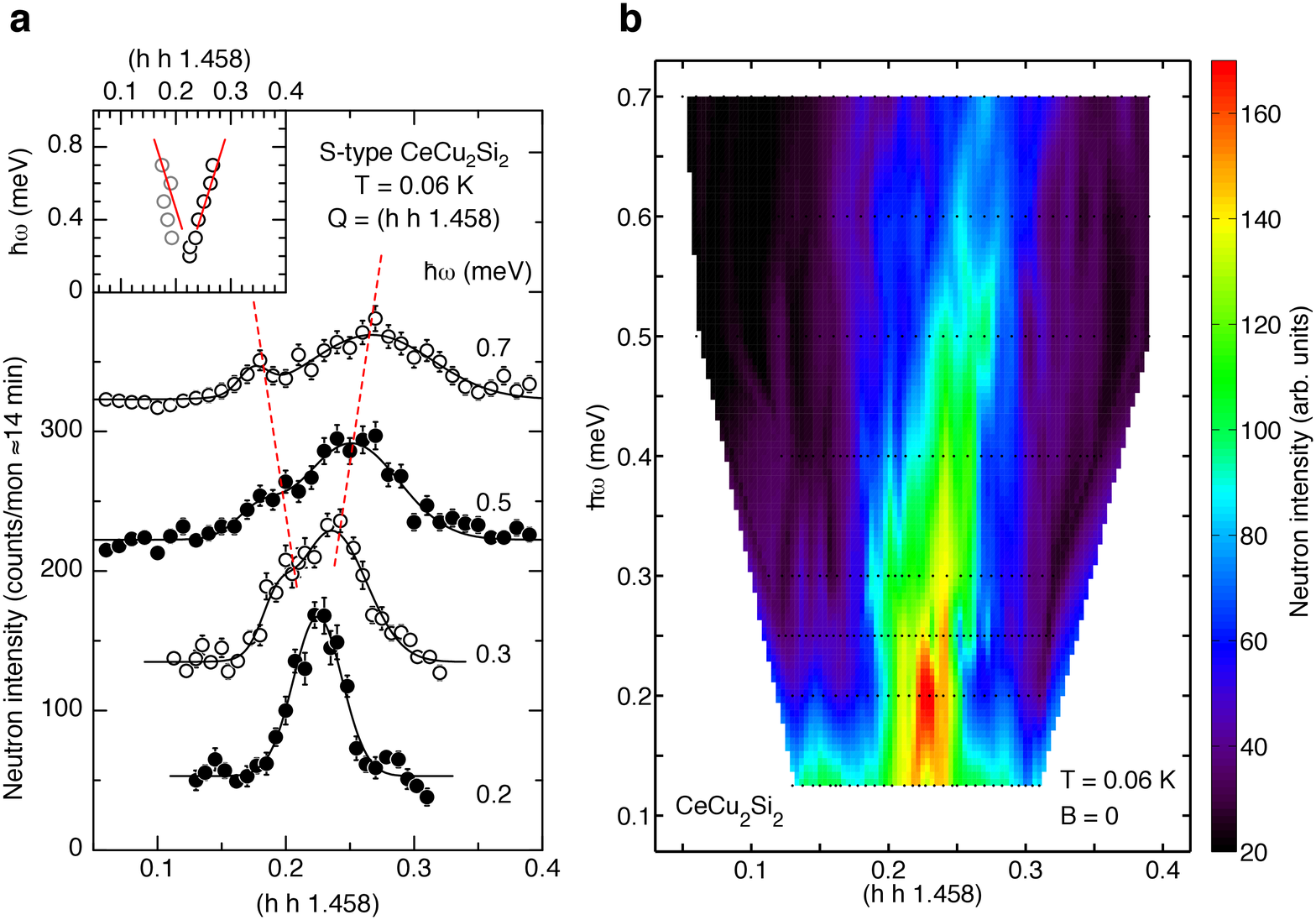}
\caption{{\bf Dispersion of the magnetic response in superconducting CeCu$_2$Si$_2$.}
(a) Wave vector ${\bf Q}$ dependence of the magnetic response around 
${\bf Q}_{\rm AF}$ in S-type CeCu$_2$Si$_2$ in the superconducting state at 
$T = 0.06$\,K for different energy transfers $\hbar\omega$. The scans are 
shifted by $100$ counts with respect to each other. Solid lines denote fits of 
two peaks with Gaussian lineshape to the data, while dashed lines are only guides
 to the eye. From the linewidth at small energy transfers a dynamic correlation length $\xi \approx 25$\,{\AA} is inferred.
Inset: Dispersion of the magnetic excitation around 
${\bf Q}_{\rm AF}$ at $T = 0.06$\,K as a result of the fits to the ${\bf Q}$ scans. 
The solid line indicates a fit to the data with a linear dispersion relation 
yielding a velocity $v_{exc} = (4.44 \pm 0.86)$\,meV{\AA} (for comparison 
spin-wave velocities in other HF metals, 
UPd$_2$Al$_3$: $v = 10-15$\,meV{\AA}~\cite{Hiess.06},
URu$_2$Si$_2$: $v \approx 45$\,meV{\AA}~\cite{Wiebe.07}). The error bars represent the
statistical error.
(b) Colour coded intensity plot of the data displayed in (a) and 
additional data, clearly indicating the dispersion of the gapped spin excitation. 
Black dots mark the $({\bf Q}, \omega)$ positions data were taken.
}\label{Figure4}
\end{figure}
We now turn to the momentum dependence of the magnetic response around
${\bf Q}_{\rm AF}$ in the superconducting state. Fig.~4 displays ${\bf Q}$ 
scans along $(h~h~1.458)$ across ${\bf Q}_{\rm AF}$ recorded at different energy
transfers $\hbar\omega$ and at $T = 0.06$\,K. 
The single peak seen at 
$\hbar\omega \approx 0.2$\,meV splits upon increasing energy transfer into
two peaks which move further apart from each other
along with a marked decrease in intensity. 
Fits with two peaks of Gaussian lineshape (solid lines) yield a good description 
of the data.
The peak positions for different $\hbar\omega$, drawn in the inset 
of Fig. 4(a), yield a linear dispersion relation. 
We conclude that the spin excitations are part of an {\it overdamped dispersive} mode.
Its velocity as read off the slope of the dispersion curve,
$v_{exc}=(4.44 \pm 0.86)$ \,meV{\AA},
is substantially smaller than the strongly renormalised Fermi velocity $v_{\rm F}^*
\approx 57$\,meV{\AA} \cite{Rauchschwalbe.82} ($1$\,meV{\AA} $ = 153$\,m/s). This indicates
a retardation of the coupling between the heavy quasiparticles and the quantum-critical
spin excitations.

\subsection*{Superconducting condensation and magnetic exchange energies}
\begin{figure}[ht]
\includegraphics[clip,width=0.9\linewidth]{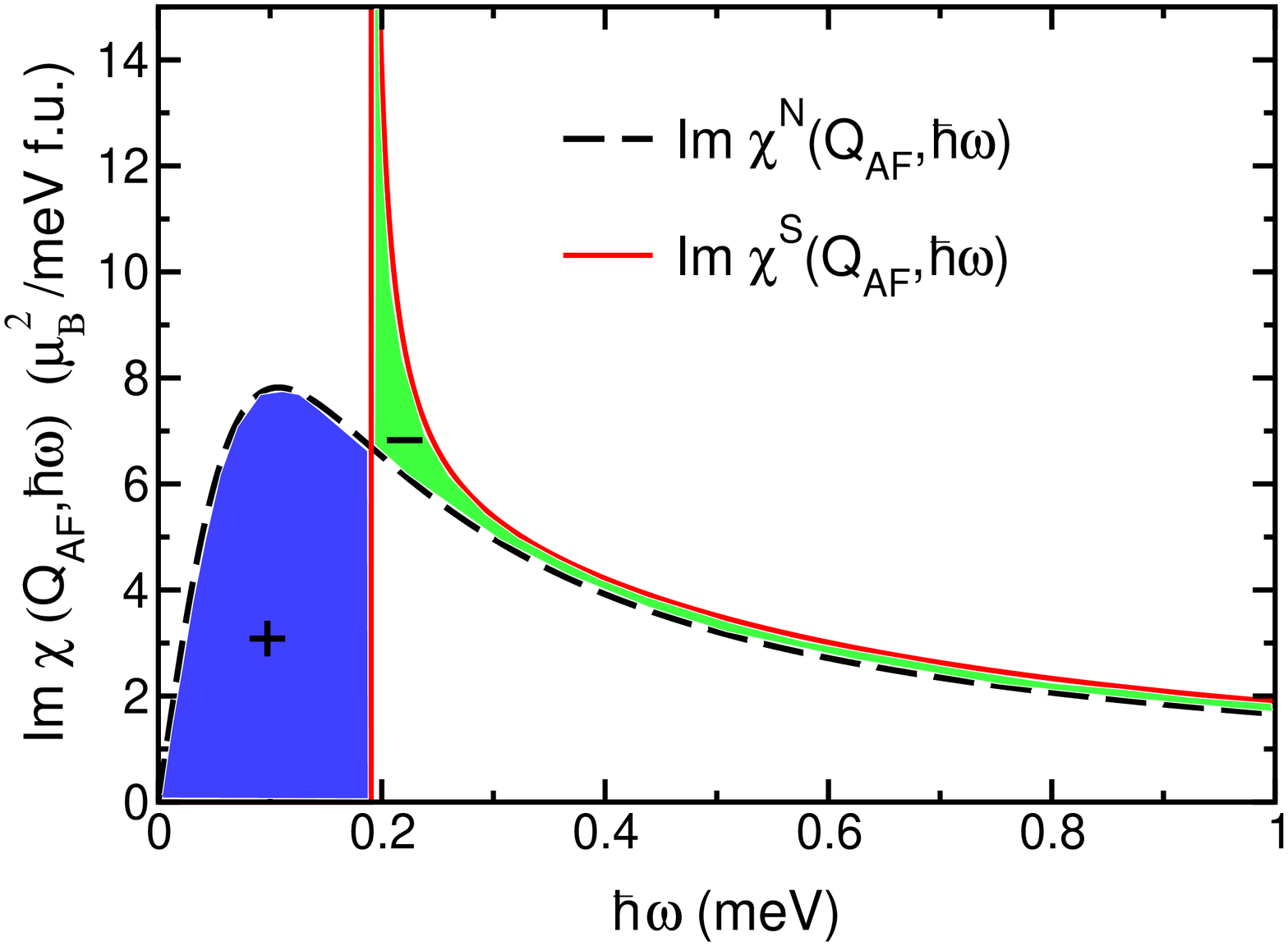}\\
\caption{{\bf
Schematic plot of the imaginary part of the dynamic spin susceptibility 
$\mbox{Im}\chi({\bf Q_{\rm AF}},\omega)$ 
in the
normal (N) and superconducting (S) states.} 
The dynamic correlation function $S({\bf Q_{\rm AF}},\omega)$ of Figure 3(a)
is related to $\mbox{Im}\chi({\bf Q_{\rm AF}},\omega)$ via the fluctuation-dissipation
theorem after de-convolving the data with the instrument's energy resolution function. The
blue area marked with a '+' contributes to an increase in $\Delta E_x$ whereas the green area
(marked with a '-') leads to a decrease in $\Delta E_x$. The fact that the opening of the gap contributes to the saving in exchange energy is a consequence
of $I({\bf Q_{\rm AF}})>0$ 
at the wave vector $\bf Q_{\rm AF}$, where  
$\mbox{Im}\chi^{N/S}({\bf Q},\omega)$ is peaked.}
\label{energygain}
\end{figure}

The observed spin excitations both below and above $T_c$ allow us to estimate 
the decrease of magnetic exchange energy in the superconducting state as compared to the putative normal state.
This saving of exchange energy 
is determined as 
follows~\cite{Scalapino.98,Leggett.98}:
\begin{eqnarray*}
\Delta E_x^{} \equiv E_x^{N}-E_x^{S}&=&
\frac{1}{g^2 \mu_B^2}\int_0^{\infty} \frac{d (\hbar\omega)}{\pi} \big 
[ n(\hbar \omega)+1 \big ] \times \nonumber \\
&& \Bigg < I({\bf q})\,
\Big[\mbox{Im}\chi^{N}({\bf q},\omega)-\mbox{Im} \chi^{S}({\bf q},\omega)\Big ]
\Bigg >,
\end{eqnarray*}
where $E_x^{N/S}$ is respectively 
the exchange energy in the normal (N) and superconducting (S) states,
$<>$ indicates an average over the first Brillouin zone, and ${\bf q} = (q_x,q_y,q_z)$ denotes a momentum transfer in the first Brillouin zone, i.e., ${\bf Q} = {\bf G} + {\bf q}$.
$I({\bf q})$ is the exchange interaction between the  localised $f$-moments
and contains nearest ($I_1$) and next nearest ($I_2$) neighbour terms:
\begin{equation}
I({\bf q}) = I_1 [cos(q_x a)+cos(q_y a)]+I_2 f_2(a,c,{\bf q})
\end{equation}
 where
$a$ and $c$ are the lattice constants, and the precise form of $f_2$ is given in~\cite{natureonline}.
The inclusion of the next nearest neighbour terms is a consequence of the three-dimensional
nature of the spin excitations of CeCu$_2$Si$_2$ (cf. Fig.~\ref{phasediagram}(c)). This is different from the cuprate superconductors and
e.g. CeCoIn$_5$, where the observed behaviour is predominantly two-dimensional. 
As described in detail in the supplementary material \cite{natureonline}, we find a magnetic exchange 
energy saving of $\Delta E_x^{} = \eta\,4.8 \cdot 10^{-3}\mbox{meV}$  per Ce ($\eta \approx 1.25$, $\eta$ being a measure of the SC volume fraction \cite{natureonline}).
This energy saving stems primarily from the 
spectrum at low energies, below the magnetic excitation gap. This follows from the fact that the spin excitations
are peaked around the wave vector ${\bf Q}_{\rm AF}$ at which $I({\bf q})$ is positive.
Fig.~\ref{energygain} illustrates which part of the spectrum of $\mbox{Im}\chi({\bf{Q}}_{\rm AF},\omega)$ increases/decreases $\Delta E_x^{}$.
This energy gain must be compared with the superconducting
condensation energy
$\Delta E_C^{}$, which  is the difference in internal energy between the 
(putative) normal and the superconducting state at $T=0$~\cite{Scalapino.98,Leggett.98}:
\begin{equation}
\Delta E_C=U_N(T=0)-U_S(T=0).
\end{equation}
Using the specific heat data shown in Fig.~6(a) \cite{natureonline}, 
we find the condensation energy
to be $\eta\,2.27\cdot 10^{-4}\mbox{meV}$ per Ce.
Compared to the high-T$_c$ cuprates where similar analyses have been performed~\cite{Demler.98,Woo.06,Dahm.09}, 
the considerably lower energy scales  in the HF systems enable us to perform a quantitative
analysis of the data in terms of an accessible putative normal state. 
As noted above, extrapolating the spin excitations from above $T_c$ is in good qualitative agreement with the excitations of the
field-driven normal state at the lowest temperatures.
Furthermore, the electronic specific heat of both the superconducting and 
normal state can be reliably determined since phonons do not contribute at such low temperatures.
Despite this apparent advantage of  HF systems, $\Delta E_C^{}$ and  $\Delta E_x^{}$ have
not received much attention in the context of HF SC. $\Delta E_C^{}$ has been determined for 
CeCoIn$_5$~\cite{Stock.08}, 
a compound whose proximity to quantum criticality is not yet certain, since SC sets in before AF order can develop. 
Our study represents the first determination of the savings in both the exchange energy and condensation energy for
a superconductor near an AF QCP, as well as for any unconventional low-temperature superconductor.

Our observation that the magnetic exchange energy saving is more than one
order of magnitude larger
than the condensation energy implies that AF excitations are the 
primary driving force for SC. A comparable factor of exchange energy saving over condensation energy
has recently been observed in the unconventional superconductor YbBa$_2$Cu$_3$O$_{6.6}$~\cite{Dahm.09}. As described above, the temperature dependence of $\omega_{\rm gap}(T)$ in CeCu$_2$Si$_2$ 
follows a rescaled BCS form.
For a conventional BCS superconductor, where $\Theta_{\rm D}\gg \omega_{\rm gap}$,
the saving in potential energy is enhanced over the condensation energy by a factor
that depends logarithmically on the ratio of Debye temperature $\Theta_{\rm D}$ and  superconducting gap $\omega_{\rm gap}(T=0)$~\cite{Haslinger.03}. 
The corresponding  enhancement factor over the condensation energy in CeCu$_2$Si$_2$,  
where the Kondo temperature $T_{\rm K}\approx 15$\,K replaces the Debye temperature $\Theta_{\rm D}$, turns out to be two.
The fact that the observed magnetic energy saving is more than a factor
20 larger than the condensation energy,
indicates a large loss in kinetic energy.
A natural origin for the latter lies in the 
Kondo effect, since the kinetic energy of the 
quasiparticles appears through the Kondo-interaction term.
Because superconducting pairing in CeCu$_2$Si$_2$ occurs
in the spin-singlet channel, the opening of the superconducting gap
therefore weakens the Kondo-singlet formation and, by extension, reduces the 
spectral weight of the Kondo resonance~\cite{natureonline}.


\subsection*{Comparison with other unconventional superconductors}
Our understanding of the magnetic exchange energy saving in the
HF superconductor CeCu$_2$Si$_2$ near its AF QCP
naturally leads us to ask whether the effect is universal.
SC-induced enhancement of the spin-fluctuation spectrum
in some frequency range has also been observed in the high-$T_c$
cuprates such as YBa$_2$Cu$_3$O$_{7-\delta}$ \cite{Sidis.04}, 
iron pnictides such as K- or Co-doped BaFe$_2$As$_2$ \cite{Christianson.08,Inosov.10} or FeTe$_{1-x}$Se$_x$ \cite{Qiu.09,Lumsden.10},
as well as two other HF compounds, 
UPd$_2$Al$_3$ \cite{Bernhoeft.98,Sato.01} and CeCoIn$_5$ \cite{Stock.08}. 
However, there are some striking differences between the spectrum
observed in CeCu$_2$Si$_2$ and those seen in the other superconductors.
In contrast to CeCu$_2$Si$_2$, where SC and long-range 
AF order exclude each other, SC in UPd$_2$Al$_3$ occurs 
inside the antiferromagnetically ordered part of its magnetic
phase diagram, which is far away from any QCP~\cite{Link.95}. Whether a QCP underlies SC
in the cuprates, the iron pnictides, or CeCoIn$_5$, is 
yet to be established. The normal state magnetic response of S-type CeCu$_2$Si$_2$ at ${\bf Q}_{\rm AF}$
slows down considerably, when lowering the temperature, indicating its proximity to a QCP, and displays pronounced dispersion.
CeCu$_2$Si$_2$ represents, therefore, the 
only system in which we can unambiguously establish the linkage
between AF quantum criticality and unconventional
SC, even though the effect may well prove to be
broadly relevant.
In comparison to other HF superconductors, the
inelastic spin response in CeCu$_2$Si$_2$ is broad in energy and extends 
beyond $10$ times the 
gap value, while in CeCoIn$_5$  a rather sharp, resolution-limited spin resonance
is found \cite{Stock.08}. Furthermore, unlike CeCoIn$_5$ the temperature
dependences of the spin excitation gap in CeCu$_2$Si$_2$ and 
UPd$_2$Al$_3$ \cite{Bernhoeft.99} do follow the expected BCS form.
In comparison to CeCu$_2$Si$_2$, UPd$_2$Al$_3$ also exhibits a dispersive spin excitation starting at the low-energy inelastic line (related to the edge of the spin gap \cite{Sato.01})
 with a slightly higher in-plane mode velocity \cite{Hiess.06}.
However, the situation in the cuprate superconductors is more complex, with 
an hour-glass like dispersion of the resonance 
mode \cite{Pailhes.04,Hayden.04,Tranquada.04}.

Experimentally the most prominent difference between  CeCu$_2$Si$_2$ and other 
unconventional superconductors
is the ${\bf Q}$ position of the spin excitation gap, 
which is observed in all reported unconventional superconductors at or close to simple commensurate 
positions with half-integer indices. 
E.g.,  in UPd$_2$Al$_3$ and CeCoIn$_5$ it occurs at commensurate positions, 
${\bf Q} = (0~0~1/2)$ and $(1/2~1/2~1/2)$ respectively \cite{Hiess.06,Stock.08}.
In contrast, in S-type CeCu$_2$Si$_2$ the gapped spin excitations are restricted to the vicinity of the ordering 
wave vector of the system, $\tau \approx (0.215~0.215~0.53)$,  which is incommensurate, far away from a simple commensurate value.
As a result, the opening of a spin gap becomes the major source of exchange energy saving.
By extension,  an additional excitonic 
resonance in $\chi^{S}({\bf Q},\omega)$ due to the superconducting state~\cite{Eremin.08} would reduce the energy saving. 
This is a striking difference 
between CeCu$_2$Si$_2$ on the one hand, and CeCoIn$_5$ \cite{Stock.08} and high-T$_c$ cuprate 
superconductors \cite{Woo.06} on the other. 

In conclusion, our inelastic neutron scattering experiments in CeCu$_2$Si$_2$ reveal
spin excitations associated with the AF (3D-SDW) QCP.
These spin excitations are overdamped, dispersive and gapped
in the superconducting state. Our quantitative
estimate of both, the change in magnetic exchange energy and the superconducting
condensation energy identifies the AF excitations
as a major driving force for SC. 
AF QCPs are currently being 
explored in a variety of strongly correlated electron systems,
including the new Fe pnictide superconductors.
Ba(Fe$_{1-x}$Co$_x$)$_2$As$_2$ \cite{Chu.09}, for instance,
exhibits a $T-x$ phase diagram very similar to the $T-p$ phase
diagram of CePd$_2$Si$_2$, raising the prospect that 
AF quantum critical excitations also drive 
the superconducting pairing in these new high-$T_c$ superconductors.



\section*{Author contributions} 
H.S.J. and C.G. synthesised the sample. O.S., J.A., E.F., M.L., K.S. and W.S. performed the measurements. O.S., J.A. and S.K. analysed the data. S.K. and Q.S. carried out theoretical calculations. O.S., S.K., Q.S. and F.S. wrote the manuscript. O.S., S.K. and F.S. planned and managed the project.

\section*{Additional information} 
The authors declare that they have no competing financial interests. Supplementary information accompanies this paper on www.nature.com/naturephysics. Reprints and permissions information is available online at http://npg.nature.com/reprintsandpermissions. Correspondence and requests for materials should be addressed to O.S.

\section*{Acknowledgements}
We greatly acknowledge helpful discussions with A. Chubukov, P. Coleman, T. Dahm, 
I. Eremin, B. F{\aa}k, A. Hiess and P. Thalmeier. This work was supported
by the Deutsche Forschungsgemeinschaft through Forschergruppe 960 "Quantum phase transitions", as well as
by the 
NSF Grant No. DMR-0424125 and the Robert A. Welch Foundation Grant
No. C-1411.

\section*{Methods}
High-resolution inelastic neutron scattering experiments were performed on the 
cold-neutron triple-axis spectrometer IN12 at the high-flux reactor of the Institut
 Laue-Langevin in Grenoble/France. A vertical focusing graphite (002) monochromator 
and a doubly focused (vertical and horizontal) graphite (002) analyzer were used. 
The horizontal collimation was given by the neutron guide in front of the monochromator
 and $60^\prime$ before the sample, while no collimation was inserted in the scattered
 beam. A liquid-nitrogen cooled Be filter was placed in the incident neutron beam to
 reduce higher-order contamination. The measurements were carried out with a fixed
 final wave vector $k_f = 1.15$\,{\AA}$^{-1}$ which corresponds to a final 
neutron energy $E_f = 2.74$\,meV and yields a high energy resolution 
$\Delta E \approx 57\,\mu$eV (FWHM, i.e., full width at half maximum). 
All experiments were performed on an S-type CeCu$_2$Si$_2$ single crystal
 ($m \approx 2$\,g). The crystal was mounted with the $[1\overline{1}0]$ 
axis vertical on a copper pin attached to the mixing chamber of a dilution
 refrigerator. The setup results in a $[110]-[001]$ scattering plane. 
Data were taken at temperatures between $T = 0.06$\,K and $1.7$\,K and in
 magnetic fields up to $B = 2.5$\,T applied along the vertical 
$[1\overline{1}0]$ axis. The inelastic neutron scattering measurements
 were converted to units of $\mu_{\rm B}^2/({\rm meV~f.u.})$ by 
normalizing the intensities to the incoherent scattering of the sample.
\renewcommand{\theequation}{S\arabic{equation}}

\newpage
\noindent
\textbf{\large \sffamily Supplementary Material for ``Magnetically driven superconductivity in \boldmath{CeCu$_2$Si$_2$}''}
\vspace{\baselineskip}
\newline
\noindent
\textbf{
O.~Stockert$^1$, 
J.~Arndt$^1${},
E.~Faulhaber$^{2,3}${},
C.~Geibel$^1${},
H.~S.~Jeevan$^{1,4}${},
S.~Kirchner$^{1,6}${},
M.~Loewenhaupt$^2$,
K.~Schmalzl$^5$,
W.~Schmidt$^5$,
Q.~Si$^7$,
F.~Steglich$^1$
}
\newline {\small
$^1$ {Max Planck Institute for Chemical Physics of Solids, N{\"o}thnitzer Str.~40, 01187 Dresden, Germany}\\
$^2$ {Institut f\"ur Festk\"orperphysik, Technische Universit\"at
  Dresden, 01062 Dresden, Germany}\\
$^3$ {Gemeinsame Forschergruppe Helmholtz-Zentrum Berlin - TU Dresden, 85747 Garching, Germany}\\
$^4$ {I. Physikalisches Institut, Universit\"at G\"ottingen, 37077 G\"ottingen, Germany}\\
$^5$ {J\"ulich Centre for Neutron Science at Institut Laue-Langevin, 38042 Grenoble, France}\\
$^6$ {Max Planck Institute for the Physics of Complex Systems, N{\"o}thnitzer Str.~38, 01187~Dresden, Germany}\\
$^7${Department of Physics and Astronomy, Rice University, Houston, Texas 77005-1892, USA}
}
\vspace{\baselineskip}

\subsection*{Sample characterization}
The present neutron scattering experiments were complemented by bulk measurements 
on the same S-type CeCu$_2$Si$_2$ single crystal. The heat capacity was recorded 
using a compensated heat-pulse technique, while the ac susceptibility was measured
 with a homemade susceptibility setup and was recorded during the neutron scattering
 experiment simultaneously while the neutron data were taken.
\begin{figure}[ht]
\includegraphics[clip,width=0.9\linewidth]{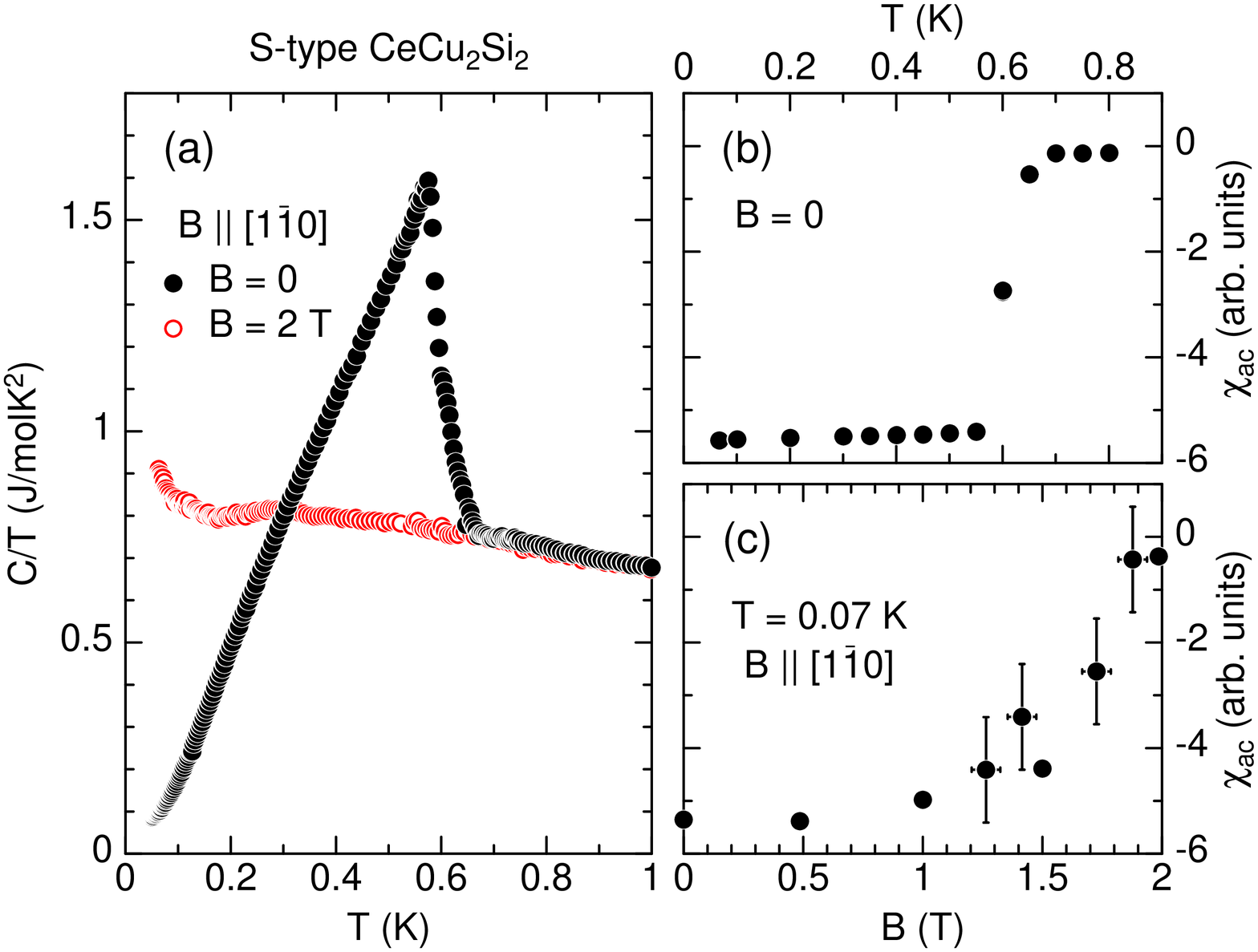}\\
\caption{
(a) Specific heat $C$ of the same S-type CeCu$_2$Si$_2$ single crystal 
which was studied by neutron scattering, plotted as $C/T$ versus temperature $T$ in 
zero magnetic field $B = 0$ and a magnetic field $B = 2$\,T applied along 
$[1\overline{1}0]$. (b) Temperature dependence of the ac susceptibility 
$\chi_{ac}$ at $B = 0$. (c) Magnetic field dependence of $\chi_{ac}$ at a 
temperature $T = 0.07$\,K and in a magnetic field $B \parallel [1\overline{1}0]$.
 The data with large error bars were taken when sweeping the magnetic field 
while all other data were measured during neutron scattering scans where the 
magnetic field was constant for a long time.}
\end{figure}
As shown in Fig. 6(a) the specific heat, plotted as $C/T$ vs. $T$, exhibits a 
pronounced maximum indicating the onset of superconductivity at $T_c = 0.6$\,K. 
The shape of the anomaly at $T_c$ in comparison to heat capacity measurements on
 other CeCu$_2$Si$_2$ samples \cite{Steglich.79,Steglich.96} and its full 
suppression already in a magnetic field of $B = 2$\,T clearly indicate that 
this is a transition into the superconducting state. The superconducting nature
 of this phase transition is further confirmed by ac susceptibility measurements
 displayed in Figs. S1(b) and (c). A sharp drop of $\chi_{ac}(T)$ at 
$T_c \approx 600$\,mK is observed with a large diamagnetic signal at lower
 temperatures. Furthermore, the susceptibility measurements yield an upper critical
 magnetic field $B_{c2}$ to kill superconductivity of $B_{c2} = 1.7$\,T 
 for $B \parallel [1\overline{1}0]$ at $T = 0.07$\,K.\\
From entropy the single-ion Kondo temperature $T_{\rm K}$ was deduced to almost coincide with
the lattice coherence temperature as obtained from the position of the low-temperature peak in the temperature dependence of the electrical resistivity \cite{Assmus.84}. Both are around 15 K. 
From the Doniach criterion together with the fact that our sample is almost quantum critical it follows that the 
magnetic energy scale $T_{\rm RKKY} \approx T_{\rm K} \approx 15$\,K.\\
\begin{figure}[ht]
\includegraphics[clip,width=0.9\linewidth]{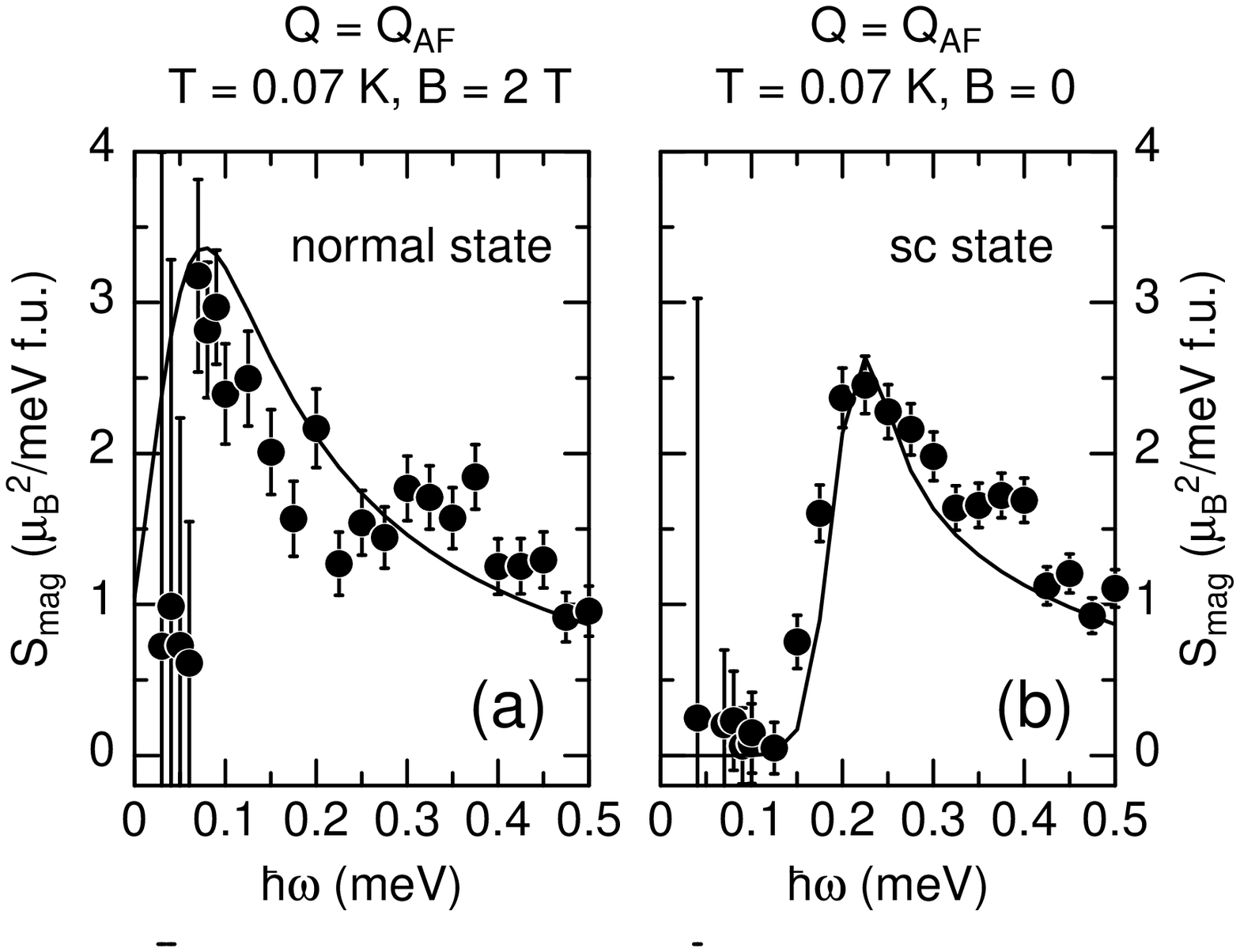}\\
\caption{Magnetic response at ${\bf Q}_{\rm AF}$ in (a) the normal and (b) the superconducting states of CeCu$_2$Si$_2$ 
(same data as in Fig.~3(a)). Solid lines are fits to the data (see text). For energies $\hbar \omega>0.5$\,meV, the dynamical susceptibility of the normal and superconducting states coincide (within the experimental error bars) and thus do not contribute to the exchange energy saving. }
\label{newfits}
\end{figure}

\subsection*{Analysis of magnetic response}
The neutron intensity is directly proportional to the scattering function 
$S({\bf Q},\omega)$. In the paramagnetic, normal state of S-type CeCu$_2$Si$_2$
 the magnetic response was fitted by a quasielastic signal with Lorentzian 
lineshape (cf. Figs. 2(a) and (b) and Fig. 3(a)). Hence, the scattering function 
$S({\bf Q},\omega)$ at momentum transfer $\hbar{\bf Q}$ and energy transfer 
$\hbar\omega$ takes the form
\begin{displaymath}
S_{qe,mag}({\bf Q},\omega) = [n(\hbar \omega)+1] \cdot \mbox{Im} \chi({\bf Q},\omega) =
[n(\hbar \omega)+1] \cdot \frac{\hbar \omega \chi_0}{\pi \Gamma} 
\cdot \frac{1}{1+(\hbar \omega /\Gamma)^2}, 
\end{displaymath}
at temperature $T$ where $n(\hbar \omega)+1 = 1/[1 - \exp(-\hbar\omega/kT)] $ is the Bose factor and  $k$ is Boltzmann constant.
Here, $\chi_0$ denotes the susceptibility and $\Gamma$ the energy linewidth of 
the fluctuations (HWHM, half width at half maximum), which is inversely 
proportional to the lifetime $\tau$ of the fluctuations. 
In the superconducting state the magnetic response was modeled using a 
modified scattering function taking the electronic density of states in a 
superconductor, $Z(\omega)$, into account, i.e.,
\begin{displaymath}
Z(\omega) = \left\{ 
\begin{array}{ll}
\frac{\displaystyle\omega}{\displaystyle\sqrt{\omega^2 - \omega_{gap}^2}} 
& \mbox{for $\omega \geq \omega_{gap}$}\\
0 & \mbox{otherwise.}\\
\end{array}
\right.
\end{displaymath}
$\omega_{gap} = \omega_{gap}(T)$ is the value of the superconducting gap at a 
certain temperature $T$.
$S_{ine,mag}({\bf Q},\omega) = S_{qe,mag}({\bf Q},\omega)\cdot Z(\omega)$ was 
then used to describe the data in the superconducting state (cf. Figs. 2(a), (b) and 3(a)).
 In each case the scattering function was convolved with the instrumental 
resolution to fit the experimental data.

\subsection*{Determination of condensation and magnetic exchange energies}
The condensation energy $\Delta E_C$
characterises the stability
of the superconducting state (S)
against a putative normal state (N) and is the difference in internal energy, or
\begin{equation}
\label{eq:Gibbs}
\lim_{T\rightarrow 0}\big(G_{N}(T,B=0)-G_{S}(T,B=0)\big)
=\mu_{0}V\int_{0}^{B_{c2}}M(T=0)\, dB=\frac{1}{2}\mu_{0}V B_{c}^{2},
\end{equation}
where $G_{S}/G_{N}$ is the Gibbs free energy of the superconducting (S)/normal (N) state
and $B_{c}^{}$ is the thermodynamic critical field
defined via $\frac{1}{2}B_{c}^{2}\equiv\int_{0}^{B_{c2}}M\, dH$. 
For CeCu$_2$Si$_2$, $B_c^{}$ has been determined 
in Ref.~\cite{Rauchschwalbe.82}.  However,  the co-existence
of small, magnetically ordered regions in the superconducting 
S-type  CeCu$_2$Si$_2$ makes it necessary to perform all energy 
estimates on the same sample.
Eq.(\ref{eq:Gibbs}) implies that $\Delta E_C$ can be obtained from the 
specific heat data of  Fig. 6(a):
\begin{eqnarray}
\label{eq:C.E.}
\Delta E_C&=&\int_{0}^{T_{c}}dT\int_{0}^{T}dT^{'}\frac{1}{T^{'}}
(C_{V}^{N}(T^{'})-C_{V}^{S}(T^{'}))\nonumber \\
&=& \eta \, 431\, \mbox{J/m}^3= \eta \, 2.27\cdot 10^{-4}\,\mbox{meV/Ce}.  
\end{eqnarray}
Here and in what follows, we take the finite field ($B=2$\,T) data as
the putative normal state. 
From the lowest temperature ($T<0.1$\,K) the data were extrapolated to $T=0$ and the contribution from the nuclear moments was subtracted.
The contribution from below $T = 0.1$\,K to $\Delta E_C$ is  tiny
($\Delta E_C^{T<0.1\,{\rm K}}=\eta \, 7.6\, \mbox{J/m}^3 = \eta \, 4\cdot 10^{-6}$\,meV/Ce). The factor 
$\eta>1$ accounts for the fact that only the
superconducting volume fraction ($1/\eta$) contributes to $\Delta E_C$, and it may 
very well be sample dependent.

We now turn to estimating the difference in  exchange energy between 
the normal and superconducting states,
in order to ascertain whether the magnetic excitations contribute 
significantly to the condensation energy.
Heavy-fermion metals are best described by the Anderson lattice model~\cite{Hewson.93}.
To estimate the magnetic exchange energy in such a system, 
it is sufficient to consider the magnetic limit of the Anderson lattice model, i.e.,
the Kondo lattice model:
\begin{equation}
H_{\small KL}=\sum_{{\bf k},\sigma} \epsilon_{\bf k} c^{\dagger}_{{\bf k}\sigma}c_{{\bf k}\sigma}^{}+J_{K}\sum_{i} {\bf S}_i
\cdot {\bf s}_c({\bf r}_i), 
\end{equation}
where ${\bf S}_i$ is the localised moment at a cerium site ${\bf r}_i$ 
which is coupled via the Kondo coupling $J_K$ to the
spd conduction electron spin density ${\bf s}_c({\bf r}_i)$ at the cerium site, and $\epsilon_{\bf k}$ describes 
the  conduction electron
bandstructure. The Kondo model describes not only the formation of heavy quasiparticles but also the AF phase and the quantum critical point, since
$H_{\small KL}$ implicitly contains the RKKY interaction among the Ce moments:
\begin{equation}
H_{\small RKKY}=\sum_{i<j} I_{i,j}  {\bf S}_i \cdot  {\bf S}_j,
\end{equation}
where the exchange constants $ I_{i,j} \sim J_K^{2}$.

We model the exchange interaction between the localised Ce-moments by including
nearest neighbour and next nearest neighbour terms appropriate for the
tetragonal, body-centred unit cell:
\begin{eqnarray}
\label{IQ}
I({\bf q}) &=& I_{1}\big [ \cos(q_xa)+\cos(q_ya) \big]
+I_{2}\big [ \cos(q_xa/2+q_ya/2+q_zc/2)
 \nonumber \\
&+&\cos(-q_xa/2+q_ya/2+q_zc/2)+\cos(-q_xa/2-q_ya/2+q_zc/2)\nonumber \\
&+&\cos(q_xa/2-q_ya/2+q_zc/2)+\cos(p_xa + p_ya) + \cos(p_ya - p_xa)\big ],
\end{eqnarray}
where 
$a=4.1$\,{\AA} and $c=9.9$\,{\AA} are the lattice constants for the $a$-
and $c$-axis, respectively and $I_{1}$ and $I_{2}$ are the nearest and next 
nearest neighbour exchange interactions.
The energy saving  in magnetic  exchange energy of the superconducting ground state compared to the putative
normal ground state is then given by
\begin{eqnarray}
\label{exchange-energy}
\Delta E_x^{}&\equiv& E_x^{N}-E_x^{S}\,=
\, \frac{A}{g^2 \mu_B^2}\int_0^{\infty} \frac{d (\hbar\omega)}{\pi} \big 
[ n(\hbar \omega)+1 \big ]  \\
&\times& \int^{\pi/a}_{-\pi/a} d q_x
\int^{\pi/a}_{-\pi/a} d q_y \int^{\pi/c}_{-\pi/c} d q_z \, \, I({q_x,q_y,q_z})
\nonumber \\
&\times&
\mbox{Im} \Big [\chi^{N}(q_x,q_y,q_z,\hbar\omega)-\chi^{S}(q_x,q_y,q_z,\hbar\omega)\Big ]
{\Bigg /}\int^{\pi/a}_{-\pi/a} d q_x
\int^{\pi/a}_{-\pi/a} d q_y \int^{\pi/c}_{-\pi/c} d q_z. \nonumber
\end{eqnarray}
Crystalline-electric-field effects split the $J=5/2$ states of the Ce$^{3+}$ ion up into
a ground-state doublet and a quasi-quartet at high energies ($> 30$\,meV) and result in g-factors $g_z\approx g_{\perp} \approx 2$,
and  an almost isotropic spin susceptibility~\cite{Goremychkin.93}.
$A$ is a constant given by $A=\eta \cdot 8\cdot 3/2$ resulting from the eight symmetry
equivalent incommensurate AF
wave vectors in the first Brillouin zone and  the fact that neutrons only
detect moments and spin fluctuations perpendicular to the actual momentum transfer.
Our sign convention in Eq. (\ref{eq:Gibbs}) implies that a positive $\Delta E_x^{}$ is equivalent to an energy saving in the superconducting state. Different energy ranges will in general contribute differently
in either decreasing or increasing the exchange energy as the system
goes from the (putative) T=0 normal state to the superconducting state.

The exchange constants $I_1$ and $I_2$ of Eq.~(\ref{IQ}) follow from three independent relations: (i) the observed dispersion of Fig. 7(a), (ii)
the fact that S-type CeCu$_2$Si$_2$ is close to quantum criticality and meets the Doniach criterion and (iii) the functional form of the RKKY interaction.
To estimate the magnitude of the exchange interaction $I_1$, we chose $v_{exc}$ from the observed dispersion of the overdamped excitations (see inset of Fig. 4(a)). 
Using the mean field expression of $v_{exc}$  for a three-dimensional,
cubic lattice with $I_2=0$, we find $I_1=0.63$\,meV. Note, that an $I_2>0$ will lead to an increase in $I_1$
for fixed dispersion, making $I_1=0.63$\,meV an estimate from below.
Alternatively, estimating $I_1$ via the Doniach criterion,
we find that $I_1 \approx 0.6$\,meV when using
$T_{\rm K}\approx 15$\,K and the fact that S-type CeCu$_2$Si$_2$ is almost quantum critical.
The ratio $I_1/I_2$ is estimated from the distance-dependence
of the RKKY interaction.
This interaction shows oscillatory behaviour with an envelope that falls off as a function of the
inverse of the distance $r$ between the Ce moments. For free electrons, this function is a simple power law  
and the period of the oscillating
function is set by $2 k_{\rm F}r$. 
In CeCu$_2$Si$_2$, the RKKY interaction is mediated by the spd conduction electrons. For a reliable estimate
of the ratio $I_1/I_2$ in CeCu$_2$Si$_2$ we combine band structure calculations for the non-magnetic La-homologue LaCu$_2$Si$_2$ of reference~\cite{Jarlborg.83} with the RKKY interaction for non-spherical Fermi surfaces obtained by L.~Roth et al.~\cite{Roth.66}.
As found by Roth et al., the RKKY interaction shows a $1/r^3$ dependence
even for non-spherical Fermi surfaces except for directions where the effective band mass diverges. We therefore set $I_1/I_2 \approx r_2^3\cos(\phi_1)/(r_1^3\cos(\phi_2))$ or $I_2 \approx 0.35 I_1 \cos(\phi_2)/\cos(\phi_1)$, 
where $r_1$ ($r_2$) is the distance between
nearest (next nearest) neighbour Ce atoms and $\cos(\phi_1)$ and $\cos(\phi_2)$ are oscillating factors which not only depend on the 
magnitude of $r_{1}/r_{2}$ but also on the direction for non-spherical Fermi surfaces. 
The oscillating factors are determined by the wave vector differences of calipering pairs of points of the Fermi surface along the
[110] for the next nearest neighbour in the basal plane and [111] directions for the  next nearest neighbour in the unit cell centre~\cite{Jarlborg.83,Roth.66}.
 
As a result, the next nearest neighbour exchange constant $I_2^{\mbox{\small basal}}=-0.08 I_1$ for two Ce moments 
in the basal plane differs from  the exchange constant $I_2^{\mbox{\small center}}=0.60 I_1$ for next nearest neighbours in 
the unit cell centres.
We are only interested in a lower estimate for the exchange energy gain and  use a simple average
$I_2=(I_2^{\mbox{\small basal}}+2I_2^{\mbox{\small centre}})/3\approx 0.38 I_1$ for the next nearest neighbour exchange constant 
in  Eq.~(\ref{IQ}). Note that there are twice as many next nearest neighbours in the unit cell centre than in the basal plane.

The volume fraction $\eta$ enters Eq.~(\ref{exchange-energy}), since the magnetically ordered
regions in S and N  yield (essentially) identical responses for $B=0$ and $B=2$\,T. 
Therefore, the actual  change in exchange energy
between S and N is larger by a factor $\eta$. As a result, the ratio between 
$\Delta E_C$ and $\Delta E_{x}$ will be independent of $\eta$. Nonetheless, an estimate of the volume fraction $\eta$ of our sample can be obtained from the ordered moment associated with the elastic magnetic response and the weak anomaly in the heat capacity yielding $\approx 0.02\,\mu_{\rm B}$/Ce. Comparing this to the ordered moment in the A-phase,  $\mu_{ord} \approx 0.1\,\mu_{\rm B}$/Ce yields $\eta \approx 1.25$.

The spin susceptibility in the normal state has been parametrized as
\begin{eqnarray}
\label{Nchi}
\mbox{Im}\chi^{N}({\bf Q},\omega)&=&\frac{\chi_0}{1+\xi^2 ({{\bf Q} - {\bf Q}_{\rm AF}})^2} 
\frac{\hbar \omega/\Gamma_q}{1+(\hbar \omega/\Gamma_q)^2} \\
&&\hspace*{-4.0cm}=\frac{\chi_0}{1+(\xi/a)^2 ( a (Q_x-Q^x_{\rm AF}))^2 +(\xi/a)^2 ( a (Q_y-Q^y_{\rm AF}))^2+ (\xi/c)^2 ( c (Q_z-Q^z_{\rm AF}))^2} 
\frac{2\hbar \omega/\Gamma}{1+4(\hbar \omega/\Gamma)^2}.\nonumber
\end{eqnarray}
In the superconducting state, the susceptibility has been modeled as
\begin{eqnarray}
\mbox{Im}\chi^{S}({\bf Q},\omega)&=&
\frac{\omega}{\sqrt{\omega^2-\omega^2_{gap}}}\mbox{Im}\chi^{N}({\bf Q},\omega),
\label{Schi}
\end{eqnarray}
for $\omega \geq\omega_{gap}$ and zero otherwise.
These expressions model very well the experimental data for $\mbox{Im}\chi^{N}({\bf Q},\omega)$ and $\mbox{Im}\chi^{S}({\bf Q},\omega)$ in the vicinity of ${\bf Q}_{\rm AF}$ at small energy transfers where $\omega_{gap}$ is independent of ${\bf Q}$. 
Away from ${\bf Q}_{\rm AF}$ and at energy transfers above 0.5 meV  $\mbox{Im}\chi^{N}({\bf Q},\omega)$ and $\mbox{Im}\chi^{S}({\bf Q},\omega)$ are identical (within the experimental error bars) and therefore do not
contribute to  $\Delta E_x^{}$.

Putting everything  together, we obtain 
$\Delta E_x^{}=\eta\, 4.8\cdot 10^{-3}$\,meV/Ce
and $\Delta E_x^{}/\Delta E_C=21.1$. The parameters in 
Eq.(\ref{Nchi}) and (\ref{Schi})
were obtained from fits to the experimental data
at $T=0.07$\,K in the superconducting ($B=0$) and normal ($B=2$\,T) state 
as $\xi=20-25$\,{\AA}, $\Gamma_N=0.11$\,meV, $\chi_0^N=15.64\,\mu_{\rm B}^2$, $\Gamma_S=0.225$\,meV, 
and $\chi_0^S=8.69\,\mu_{\rm B}^2$.
We  checked that the associated static structure factors integrated over the full Brillouin zone (``local moment sum rule'')
in the normal and superconducting state yield identical results (within a 5\% error).
The (in reciprocal space) isotropic fit to $\mbox{Im}\chi^{N}$ ($\mbox{Im}\chi^{S}$), see \cite{Stockert.04},
suggests that the value for $\Delta E_x^{}$ is an estimate from below.
We also checked that  even if  $I_2$ were zero,
$\Delta E_x^{}$  would still be one order of magnitude larger than $\Delta E_C$.

A realistic modelling of the dynamic susceptibility of the normal and superconducting states in the entire momentum and 
frequency range should include the overdamped, dispersive excitations. This can be accomplished by parameterising the normal state
susceptibility by the SDW form
\begin{equation}
\label{Fit2}
\chi ({\bf Q},\omega)=\frac{\chi_0}{1 + \xi^2({\bf Q}-{\bf Q_{\rm AF}})^2 - b^2 \xi^2\omega^2 -i \xi^2 \omega/\Gamma }
\end{equation}
which includes an overdamped excitation at
\begin{equation}
\label{dispersion}
\omega = \omega_{exc}=\pm b^{-1}\sqrt{\xi^{-2}+({\bf Q}-{\bf Q_{\rm AF}})^2}
\end{equation}
where $\xi$ is the correlation length and 
$b=v^{-1}_{exc}$ follows from the dispersion relation  of Fig. 4(a).
Eq.~(\ref{Fit2}) reproduces the observed Lorentzian lineshape of the overdamped, dispersive mode for a fixed $\omega$ and
${\bf Q}\sim {\bf Q}_{exc}$ (where ${\bf Q}_{exc}$ is a solution to Eq.~(\ref{dispersion})):
\begin{equation}
\label{Lorentz}
\mbox{Im}\chi^{N}({\bf Q},\omega)=\frac{1}{\pi}\frac{\tilde{\chi}_0 \omega_{}/\tilde{\Gamma}}{({\bf Q}-{\bf Q}_{exc})^2+\omega^2_{}/\tilde{\Gamma}^2},
\end{equation}
where $\tilde{\Gamma}=2\Gamma |{\bf Q}_{AF}-{\bf Q}_{exc}|$ and $\tilde{\chi}_0=\pi/(2 \xi^2 |{\bf Q}_{AF}-{\bf Q}_{exc}|)$. Fitting the ${\bf Q}$-dependence of the data at different energies gives rise to the same width, as required by Eq.~(\ref{Lorentz}).
$\mbox{Im}\chi^{N}({\bf Q},\omega)$ and $\mbox{Im}\chi^{S}({\bf Q},\omega)$ are identical at energy transfers above 0.5 meV
and are still sizeable in magnitude for ${\bf Q}$ sufficiently away from ${\bf Q}_{AF}$. 
We found that the best fit to the superconducting $\mbox{Im}\chi^{S}$ could be obtained 
from
\begin{eqnarray}
\mbox{Im}\chi^{S}({\bf Q},\omega)&=&\Theta(\omega_{gap}-\omega)
\frac{\omega}{(\omega^3-\omega^3_{gap})^{1/3}}\mbox{Im}\chi^{N}({\bf Q},\omega).
\end{eqnarray}
Taking $\chi_0^N=7.0\,\mu_{\rm B}^2$, $\Gamma^N=25$\,meV{\AA}$^2$ and  $\chi_0^S=6.9\,\mu_{\rm B}^2$, $\Gamma^S=25$\,meV{\AA}$^2$ reproduces all features of our data (cf. Fig.~\ref{newfits}) and results in a saving in exchange energy $\Delta E_x=\eta\,4.7 \cdot 10^{-3}$\,meV/Ce. 
The slightly modified parameters $\chi_0^N=10.0\,\mu_{\rm B}^2$, $\Gamma^N=20.0$\,meV{\AA}$^2$ and  $\chi_0^S=10.0\,\mu_{\rm B}^2$, $\Gamma^S=20.0$\,meV{\AA}$^2$ capture the broad features in $\mbox{Im}\chi({\bf Q},\omega)$ and result in a fit of somewhat lesser quality 
to the data, but still yield  $\Delta E_x=\eta\,4.1 \cdot 10^{-3}$\,meV/Ce.
We therefore conclude that $\Delta E_x\gg \Delta E_C$ is a stable observation insensitive to the details of the fitting
and results primarily 
from the changes of $\mbox{Im}\chi({\bf Q},\omega)$ at low energy transfers in the vicinity of ${\bf Q}_{\rm AF}$ as superconductivity sets in.
The saving in exchange energy is more than an order of magnitude larger than the condensation energy, 
thus identifying the build up of magnetic correlations near the AF QCP 
as the major driving force for SC in CeCu$_2$Si$_2$. It is important to note, that an increase in $\Delta E_x$ comes from the opening of the spin gap and 
not the 'resonance'-like feature above the spin gap, which tends to reduce the energy saving as illustrated in Fig. 5.
This figure shows the difference
of Im $\chi^{S}$ and Im $\chi^{N}$. The blue area marked with a '+' contributes to an increase in $\Delta E_x$ whereas the green area leads to a reduction in $\Delta E_x$, as a consequence of $I({\bf Q}_{\rm AF})>0$ and similar for the other wave vectors as follows from Eq. (\ref{Nchi}). 
For e.g. CeCoIn$_5$\cite{Stock.08,Eremin.08}, the shifted spectral weight in the superconducting state has to be such that the resulting positive (since $I({\bf Q})>0$) green area exceeds the
blue area for an overall saving in exchange energy. Therefore, the spin resonance at low energies observed in CeCoIn$_5$ and in the cuprates contributes to an increase in $\Delta E_x$. A sharp resonance occurs in response to the superconducting state in predominantly two-dimensional superconductors and adds to a saving in exchange energy. It is however not expected in CeCu$_2$Si$_2$, which is a 3D superconductor as deduced from the nearly
isotropic upper critical field~\cite{Assmus.84}.
Instead, as shown in e.g. Figure 2(a), the magnetic response in CeCu$_2$Si$_2$ is broad and extends to more than ten times the gap energy in 
contrast to e.g. CeCoIn$_5$ or the cuprates where a sharp spin resonance has been observed.

Our results imply that there is a sizeable ``kinetic'' energy loss in CeCu$_2$Si$_2$.
As described in the main text, superconductivity in CeCu$_2$Si$_2$ occurs
in the spin-singlet channel. As a result of the opening of the superconducting gap,
the Kondo-singlet formation is weakened and  the  spectral weight of the Kondo resonance is reduced. The spectral weight sum rule dictates that the integration of the single-electron density of states over all energies is unchanged (and is equal to one). As a result,
the lost spectral weight from low energies (within the scale of the Kondo temperature)
must be transferred to higher energies, on the order of the on-site Coulomb 
interactions among the $f$-electrons where the incoherent $f$-electron
excitations reside. 
This energy loss in a heavy fermion superconductor should therefore be distinguished from the ordinary kinetic energy loss of a 
classical superconductor~\cite{KirchnerandSi}.


\begin{thebibliography}{10}
\expandafter\ifx\csname url\endcsname\relax
  \def\url#1{\texttt{#1}}\fi
\expandafter\ifx\csname urlprefix\endcsname\relax\def\urlprefix{URL }\fi
\providecommand{\bibinfo}[2]{#2}
\providecommand{\eprint}[2][]{\url{#2}}

\bibitem{Steglich.79}
\bibinfo{author}{Steglich, F.} \emph{et~al.}
\newblock \bibinfo{title}{Superconductivity in the presence of strong {P}auli
  paramagnetism: {CeCu$_2$Si$_2$}}.
\newblock \emph{\bibinfo{journal}{Phys.~Rev.~Lett.}}
  \textbf{\bibinfo{volume}{43}}, \bibinfo{pages}{1892--1896} (\bibinfo{year}{1979}).

\bibitem{Miyake.86}
\bibinfo{author}{Miyake, K.}, \bibinfo{author}{Schmitt-Rink, S.} \&
  \bibinfo{author}{Varma, C.~M.}
\newblock \bibinfo{title}{Spin-fluctuation-mediated even-parity pairing in
  heavy-fermion superconductors}.
\newblock \emph{\bibinfo{journal}{Phys.~Rev.~B}} \textbf{\bibinfo{volume}{34}},
  \bibinfo{pages}{6554--6556} (\bibinfo{year}{1986}).

\bibitem{Scalapino.86}
\bibinfo{author}{Scalapino, D.~J.}, \bibinfo{author}{Loh, E.} \&
  \bibinfo{author}{Hirsch, J.~E.}
\newblock \bibinfo{title}{d-wave pairing near a spin-density-wave instability}.
\newblock \emph{\bibinfo{journal}{Phys.~Rev.~B}} \textbf{\bibinfo{volume}{34}},
  \bibinfo{pages}{8190--8192} (\bibinfo{year}{1986}).

\bibitem{Monthoux.07}
\bibinfo{author}{Monthoux, P.}, \bibinfo{author}{Pines, D.} \&
  \bibinfo{author}{Lonzarich, G.~G.}
\newblock \bibinfo{title}{Superconductivity without phonons}.
\newblock \emph{\bibinfo{journal}{Nature}} \textbf{\bibinfo{volume}{450}},
  \bibinfo{pages}{1177-1183} (\bibinfo{year}{2007}).

\bibitem{Mathur.98}
\bibinfo{author}{Mathur, N.~D.} \emph{et~al.}
\newblock \bibinfo{title}{Magnetically mediated superconductivity in heavy
  fermion compounds}.
\newblock \emph{\bibinfo{journal}{Nature}} \textbf{\bibinfo{volume}{394}},
  \bibinfo{pages}{39--43} (\bibinfo{year}{1998}).

\bibitem{Gegenwart.08}
\bibinfo{author}{Gegenwart, P.}, \bibinfo{author}{Si, Q.} \&
  \bibinfo{author}{Steglich, F.}
\newblock \bibinfo{title}{Quantum criticality in heavy-fermion metals}.
\newblock \emph{\bibinfo{journal}{Nat.~Phys.}} \textbf{\bibinfo{volume}{4}},
  \bibinfo{pages}{186--197} (\bibinfo{year}{2008}).

\bibitem{Steglich.01}
\bibinfo{author}{Steglich, F.} \emph{et~al.}
\newblock \emph{\bibinfo{title}{More is different -- Fifty years of condensed
  matter physics}}, chap. \bibinfo{chapter}{Superconductivity and magnetism in
  heavy-fermions}, \bibinfo{pages}{p. 191--210} (\bibinfo{publisher}{Princeton
  University Press}, \bibinfo{year}{2001}).

\bibitem{Gegenwart.98}
\bibinfo{author}{Gegenwart, P.} \emph{et~al.}
\newblock \bibinfo{title}{Breakup of heavy fermions on the brink of {``Phase
  $A$" in CeCu$_2$Si$_2$}}.
\newblock \emph{\bibinfo{journal}{Phys.~Rev.~Lett.}}
  \textbf{\bibinfo{volume}{81}}, \bibinfo{pages}{1501--1504} (\bibinfo{year}{1998}).

\bibitem{Yuan.03}
\bibinfo{author}{Yuan, H.~Q.} \emph{et~al.}
\newblock \bibinfo{title}{Observation of two distinct superconducting phases in
  {CeCu$_2$Si$_2$}}.
\newblock \emph{\bibinfo{journal}{Science}} \textbf{\bibinfo{volume}{302}},
  \bibinfo{pages}{2104--2107} (\bibinfo{year}{2003}).

\bibitem{Rosch.99}
\bibinfo{author}{Rosch, A.}
\newblock \bibinfo{title}{Interplay of disorder and spin fluctuations in the
  resistivity near a quantum critical point}.
\newblock \emph{\bibinfo{journal}{Phys. Rev. Lett.}}
  \textbf{\bibinfo{volume}{82}}, \bibinfo{pages}{4280--4283} (\bibinfo{year}{1999}).

\bibitem{Stockert.04}
\bibinfo{author}{Stockert, O.} \emph{et~al.}
\newblock \bibinfo{title}{Nature of the {A} phase in {CeCu$_2$Si$_2$}}.
\newblock \emph{\bibinfo{journal}{Phys.~Rev.~Lett.}}
  \textbf{\bibinfo{volume}{92}}, \bibinfo{pages}{136401}
  (\bibinfo{year}{2004}).

\bibitem{Holmes.04}
\bibinfo{author}{Holmes, A.~T.}, \bibinfo{author}{Jaccard, D.} \&
  \bibinfo{author}{Miyake, K.}
\newblock \bibinfo{title}{Signatures of valence fluctuations in
  {CeCu$_2$Si$_2$} under high pressure}.
\newblock \emph{\bibinfo{journal}{Phys.~Rev.~B}} \textbf{\bibinfo{volume}{69}},
  \bibinfo{pages}{024508} (\bibinfo{year}{2004}).

\bibitem{Steglich.96}
\bibinfo{author}{Steglich, F.} \emph{et~al.}
\newblock \bibinfo{title}{New observations concerning magnetism and
  superconductivity in heavy-fermion metals}.
\newblock \emph{\bibinfo{journal}{Physica B}}
  \textbf{\bibinfo{volume}{223-224}}, \bibinfo{pages}{1--8}
  (\bibinfo{year}{1996}).

\bibitem{natureonline}
\bibinfo{howpublished}{See supplementary material on Nature physics online}.

\bibitem{Stockert.08}
\bibinfo{author}{Stockert, O.} \emph{et~al.}
\newblock \bibinfo{title}{Magnetism and superconductivity in the heavy-fermion
  compound {CeCu$_2$Si$_2$} studied by neutron scattering}.
\newblock \emph{\bibinfo{journal}{Physica B}} \textbf{\bibinfo{volume}{403}},
  \bibinfo{pages}{973--976} (\bibinfo{year}{2008}).

\bibitem{Rauchschwalbe.82}
\bibinfo{author}{Rauchschwalbe, U.} \emph{et~al.}
\newblock \bibinfo{title}{Critical fields of the "heavy-fermion" superconductor
  {CeCu$_2$Si$_2$}}.
\newblock \emph{\bibinfo{journal}{Phys.~Rev.~Lett.}}
  \textbf{\bibinfo{volume}{49}}, \bibinfo{pages}{1448--1451} (\bibinfo{year}{1982}).

\bibitem{Ohkawa.87}
\bibinfo{author}{Ohkawa, F.}
\newblock \bibinfo{title}{Cooper pairs of $d_{x^2-y^2}$-symmetry in simple
  square lattices}.
\newblock \emph{\bibinfo{journal}{J. Phys. Soc. Jpn.}}
  \textbf{\bibinfo{volume}{56}}, \bibinfo{pages}{2267--2270} (\bibinfo{year}{1987}).

\bibitem{Ishida.99}
\bibinfo{author}{Ishida, K.} \emph{et~al.}
\newblock \bibinfo{title}{Evolution from magnetism to unconventional
  superconductivity in a series of {Ce$_x$Cu$_2$Si$_2$} compounds probed by
  {C}u {NQR}}.
\newblock \emph{\bibinfo{journal}{Phys.~Rev.~Lett.}}
  \textbf{\bibinfo{volume}{82}}, \bibinfo{pages}{5353--5356} (\bibinfo{year}{1999}).

\bibitem{Fujiwara.08}
\bibinfo{author}{Fujiwara, K.} \emph{et~al.}
\newblock \bibinfo{title}{High pressure {NQR} measurement in {CeCu$_2$Si$_2$}
  up to sudden disappearance of superconductivity}.
\newblock \emph{\bibinfo{journal}{J.~Phys.~Soc.~Jpn.}}
  \textbf{\bibinfo{volume}{77}}, \bibinfo{pages}{123711}
  (\bibinfo{year}{2008}).

\bibitem{Stockert.06a}
\bibinfo{author}{Stockert, O.} \emph{et~al.}
\newblock \bibinfo{title}{Peculiarities of the antiferromagnetism in
  {CeCu$_2$Si$_2$}}.
\newblock \emph{\bibinfo{journal}{J. Phys.: Conf. Ser.}}
  \textbf{\bibinfo{volume}{51}}, \bibinfo{pages}{211--218} (\bibinfo{year}{2006}).

\bibitem{Scalapino.98}
\bibinfo{author}{Scalapino, D.~J.} \& \bibinfo{author}{White, S.~R.}
\newblock \bibinfo{title}{Superconducting condensation energy and an
  antiferromagnetic exchange-based pairing mechanism}.
\newblock \emph{\bibinfo{journal}{Phys.~Rev.~B}} \textbf{\bibinfo{volume}{58}},
  \bibinfo{pages}{8222--8224} (\bibinfo{year}{1998}).

\bibitem{Leggett.98}
\bibinfo{author}{Leggett, A.}
\newblock \bibinfo{title}{Where is the energy saved in cuprate
  superconductivity?}
\newblock \emph{\bibinfo{journal}{J.~Phys.~Chem.~Solids}}
  \textbf{\bibinfo{volume}{59}}, \bibinfo{pages}{1729--1732} (\bibinfo{year}{1998}).

\bibitem{Demler.98}
\bibinfo{author}{Demler, E.} \& \bibinfo{author}{Zhang, S.-C.}
\newblock \bibinfo{title}{Quantitative test of a microscopic mechanism of
  high-temperature superconductivity}.
\newblock \emph{\bibinfo{journal}{Nature}} \textbf{\bibinfo{volume}{396}},
  \bibinfo{pages}{733--735} (\bibinfo{year}{1998}).

\bibitem{Woo.06}
\bibinfo{author}{Woo, H.} \emph{et~al.}
\newblock \bibinfo{title}{Magnetic energy change available to superconducting
  condensation in optimally doped {YBa$_2$Cu$_3$O$_{6.95}$}}.
\newblock \emph{\bibinfo{journal}{Nat.~Phys.}} \textbf{\bibinfo{volume}{2}},
  \bibinfo{pages}{600--604} (\bibinfo{year}{2006}).

\bibitem{Dahm.09}
\bibinfo{author}{Dahm, T.} \emph{et~al.}
\newblock \bibinfo{title}{Strength of the spin-fluctuation-mediated pairing
  interaction in a high-temperature superconductor}.
\newblock \emph{\bibinfo{journal}{Nat.~Phys.}} \textbf{\bibinfo{volume}{5}},
  \bibinfo{pages}{217--221} (\bibinfo{year}{2009}).

\bibitem{Stock.08}
\bibinfo{author}{Stock, C.}, \bibinfo{author}{Broholm, C.},
  \bibinfo{author}{Hudis, J.}, \bibinfo{author}{Kang, H.~J.} \&
  \bibinfo{author}{Petrovic, C.}
\newblock \bibinfo{title}{Spin resonance in the d-wave superconductor
  {CeCoIn$_5$}}.
\newblock \emph{\bibinfo{journal}{Phys.~Rev.~Lett.}}
  \textbf{\bibinfo{volume}{100}}, \bibinfo{pages}{087001}
  (\bibinfo{year}{2008}).

\bibitem{Haslinger.03}
\bibinfo{author}{Haslinger, R.} \& \bibinfo{author}{Chubukov, A.~V.}
\newblock \bibinfo{title}{Condensation energy in strongly coupled
  superconductors}.
\newblock \emph{\bibinfo{journal}{Phys.~Rev.~B}} \textbf{\bibinfo{volume}{68}},
  \bibinfo{pages}{214508} (\bibinfo{year}{2003}).

\bibitem{Sidis.04}
\bibinfo{author}{Sidis, Y.} \emph{et~al.}
\newblock \bibinfo{title}{Magnetic resonant excitations in high-{$T_c$}
  superconductors}.
\newblock \emph{\bibinfo{journal}{phys. status solidi b}}
  \textbf{\bibinfo{volume}{241}}, \bibinfo{pages}{1204--1210} (\bibinfo{year}{2004}).

\bibitem{Christianson.08}
\bibinfo{author}{Christianson, A.~D.} \emph{et~al.}
\newblock \bibinfo{title}{Unconventional superconductivity in
  {Ba$_{0.6}$K$_{0.4}$Fe$_2$As$_2$} from inelastic neutron scattering}.
\newblock \emph{\bibinfo{journal}{Nature}} \textbf{\bibinfo{volume}{456}},
  \bibinfo{pages}{930--932} (\bibinfo{year}{2008}).

\bibitem{Inosov.10}
\bibinfo{author}{Inosov, D.~S.} \emph{et~al.}
\newblock \bibinfo{title}{Normal-state spin dynamics and temperature-dependent
  spin-resonance energy in optimally doped
  {B}a{F}e$_{1.85}${C}o$_{0.15}${A}s$_2$}.
\newblock \emph{\bibinfo{journal}{Nat. Phys.}} \textbf{\bibinfo{volume}{6}},
  \bibinfo{pages}{178--181} (\bibinfo{year}{2010}).

\bibitem{Qiu.09}
\bibinfo{author}{Qiu, Y.} \emph{et~al.}
\newblock \bibinfo{title}{Spin gap and resonance at the nesting wave vector in
  superconducting {F}e{S}e$_{0.4}${T}e$_{0.6}$}.
\newblock \emph{\bibinfo{journal}{Phys.~Rev.~Lett.}}
  \textbf{\bibinfo{volume}{103}}, \bibinfo{pages}{067008}
  (\bibinfo{year}{2009}).

\bibitem{Lumsden.10}
\bibinfo{author}{Lumsden, M.~D.} \emph{et~al.}
\newblock \bibinfo{title}{Evolution of spin excitations into the
  superconducting state in {FeTe$_{1-x}$Se$_x$}}.
\newblock \emph{\bibinfo{journal}{Nat. Phys.}} \textbf{\bibinfo{volume}{6}}, 
\bibinfo{pages}{182--186}  (\bibinfo{year}{2010}).

\bibitem{Bernhoeft.98}
\bibinfo{author}{Bernhoeft, N.} \emph{et~al.}
\newblock \bibinfo{title}{Enhancement of magnetic fluctuations on passing below
  {$T_c$} in the heavy fermion superconductor {UPd$_2$Al$_3$}}.
\newblock \emph{\bibinfo{journal}{Phys.~Rev.~Lett.}}
  \textbf{\bibinfo{volume}{81}}, \bibinfo{pages}{4244--4247} (\bibinfo{year}{1998}).

\bibitem{Sato.01}
\bibinfo{author}{Sato, N.~K.} \emph{et~al.}
\newblock \bibinfo{title}{Strong coupling between local moments and
  superconducting 'heavy' electrons in {UPd$_2$Al$_3$}}.
\newblock \emph{\bibinfo{journal}{Nature}} \textbf{\bibinfo{volume}{410}},
  \bibinfo{pages}{340--343} (\bibinfo{year}{2001}).

\bibitem{Link.95}
\bibinfo{author}{Link, P.}, \bibinfo{author}{Jaccard, D.},
  \bibinfo{author}{Geibel, C.}, \bibinfo{author}{Wassilew, C.} \&
  \bibinfo{author}{Steglich, F.}
\newblock \bibinfo{title}{The heavy-fermion superconductor {UPd$_2$Al$_3$} at
  very high pressure}.
\newblock \emph{\bibinfo{journal}{J.~Phys.: Condens.~Matter}}
  \textbf{\bibinfo{volume}{7}}, \bibinfo{pages}{373--378} (\bibinfo{year}{1995}).

\bibitem{Bernhoeft.99}
\bibinfo{author}{Bernhoeft, N.} \emph{et~al.}
\newblock \bibinfo{title}{Magnetic fluctuations above and below {$T_c$} in the
  heavy fermion superconductor {UPd$_2$Al$_3$}}.
\newblock \emph{\bibinfo{journal}{Physica B}}
  \textbf{\bibinfo{volume}{259-261}}, \bibinfo{pages}{614--620}
  (\bibinfo{year}{1999}).

\bibitem{Hiess.06}
\bibinfo{author}{Hiess, A.} \emph{et~al.}
\newblock \bibinfo{title}{Magnetization dynamics in the normal and
  superconducting phases of {UPd$_2$Al$_3$}: {I.} surveys in reciprocal space
  using neutron inelastic scattering}.
\newblock \emph{\bibinfo{journal}{J. Phys.: Condens. Matter}}
  \textbf{\bibinfo{volume}{18}}, \bibinfo{pages}{R437--R451} (\bibinfo{year}{2006}).

\bibitem{Pailhes.04}
\bibinfo{author}{Pailh{\`e}s, S.} \emph{et~al.}
\newblock \bibinfo{title}{Resonant magnetic excitations at high energy in
  superconducting {YBa$_2$Cu$_3$O$_{6.85}$}}.
\newblock \emph{\bibinfo{journal}{Phys.~Rev.~Lett.}}
  \textbf{\bibinfo{volume}{93}}, \bibinfo{pages}{167001}
  (\bibinfo{year}{2004}).

\bibitem{Hayden.04}
\bibinfo{author}{Hayden, S.~M.}, \bibinfo{author}{Mook, H.~A.},
  \bibinfo{author}{Dai, P.}, \bibinfo{author}{Perring, T.~G.} \&
  \bibinfo{author}{Dogan, F.}
\newblock \bibinfo{title}{The structure of the high-energy spin excitations in
  a high-transition-temperature superconductor}.
\newblock \emph{\bibinfo{journal}{Nature}} \textbf{\bibinfo{volume}{429}},
  \bibinfo{pages}{531--534} (\bibinfo{year}{2004}).

\bibitem{Tranquada.04}
\bibinfo{author}{Tranquada, J.~M.} \emph{et~al.}
\newblock \bibinfo{title}{Quantum magnetic excitations from stripes in copper
  oxide superconductors}.
\newblock \emph{\bibinfo{journal}{Nature}} \textbf{\bibinfo{volume}{429}},
  \bibinfo{pages}{534--538} (\bibinfo{year}{2004}).

\bibitem{Eremin.08}
\bibinfo{author}{Eremin, I.}, \bibinfo{author}{Zwicknagl, G.},
  \bibinfo{author}{Thalmeier, P.} \& \bibinfo{author}{Fulde, P.}
\newblock \bibinfo{title}{Feedback spin resonance in superconducting
  {CeCu$_2$Si$_2$ and CeCoIn$_5$}}.
\newblock \emph{\bibinfo{journal}{Phys.~Rev.~Lett.}}
  \textbf{\bibinfo{volume}{101}}, \bibinfo{pages}{187001}
  (\bibinfo{year}{2008}).

\bibitem{Chu.09}
\bibinfo{author}{Chu, J.-H.}, \bibinfo{author}{Analytis, J.~G.},
  \bibinfo{author}{Kucharczyk, C.} \& \bibinfo{author}{Fisher, I.~R.}
\newblock \bibinfo{title}{Determination of the phase diagram of the
  electron-doped superconductor {Ba(Fe$_{1-x}$Co$_x$)$_2$As$_2$}}.
\newblock \emph{\bibinfo{journal}{Phys.~Rev.~B}} \textbf{\bibinfo{volume}{79}},
  \bibinfo{pages}{014506} (\bibinfo{year}{2009}).

\bibitem{Wiebe.07}
\bibinfo{author}{Wiebe, C.~R.} \emph{et~al.}
\newblock \bibinfo{title}{Gapped itinerant spin excitations account for missing
  entropy in the hidden-order state of {URu$_2$Si$_2$}}.
\newblock \emph{\bibinfo{journal}{Nat.~Phys.}} \textbf{\bibinfo{volume}{3}},
  \bibinfo{pages}{96--99} (\bibinfo{year}{2007}).

\end{thebibliography}

\begin{thebibliography}{10}
\expandafter\ifx\csname url\endcsname\relax
  \def\url#1{\texttt{#1}}\fi
\expandafter\ifx\csname urlprefix\endcsname\relax\def\urlprefix{URL }\fi
\providecommand{\bibinfo}[2]{#2}
\providecommand{\eprint}[2][]{\url{#2}}

\bibitem{Steglich.79}
\bibinfo{author}{Steglich, F.} \emph{et~al.}
\newblock \bibinfo{title}{Superconductivity in the presence of strong {P}auli
  paramagnetism: {CeCu$_2$Si$_2$}}.
\newblock \emph{\bibinfo{journal}{Phys.~Rev.~Lett.}}
  \textbf{\bibinfo{volume}{43}}, \bibinfo{pages}{1892--1896} (\bibinfo{year}{1979}).

\bibitem{Steglich.96}
\bibinfo{author}{Steglich, F.} \emph{et~al.}
\newblock \bibinfo{title}{New observations concerning magnetism and
  superconductivity in heavy-fermion metals}.
\newblock \emph{\bibinfo{journal}{Physica B}}
  \textbf{\bibinfo{volume}{223-224}}, \bibinfo{pages}{1--8}
  (\bibinfo{year}{1996}).

\bibitem{Assmus.84}
\bibinfo{author}{Assmus, W.} \emph{et~al.}
\newblock \bibinfo{title}{Superconductivity in {CeCu$_2$Si$_2$} single
  crystals}.
\newblock \emph{\bibinfo{journal}{Phys.~Rev.~Lett.}}
  \textbf{\bibinfo{volume}{52}}, \bibinfo{pages}{469--472} (\bibinfo{year}{1984}).

\bibitem{Rauchschwalbe.82}
\bibinfo{author}{Rauchschwalbe, U.} \emph{et~al.}
\newblock \bibinfo{title}{Critical fields of the "heavy-fermion" superconductor
  {CeCu$_2$Si$_2$}}.
\newblock \emph{\bibinfo{journal}{Phys.~Rev.~Lett.}}
  \textbf{\bibinfo{volume}{49}}, \bibinfo{pages}{1448--1451} (\bibinfo{year}{1982}).

\bibitem{Hewson.93}
\bibinfo{author}{Hewson, A.~C.}
\newblock \emph{\bibinfo{title}{{The Kondo problem to Heavy Fermions}}}
  (\bibinfo{publisher}{Cambridge University Press, Cambridge},
  \bibinfo{year}{1993}).

\bibitem{Goremychkin.93}
\bibinfo{author}{Goremychkin, E.~A.} \& \bibinfo{author}{Osborn, R.}
\newblock \bibinfo{title}{Crystal-field excitations in {CeCu$_2$Si$_2$}}.
\newblock \emph{\bibinfo{journal}{Phys.~Rev.~B}} \textbf{\bibinfo{volume}{47}},
  \bibinfo{pages}{14280--14290} (\bibinfo{year}{1993}).

\bibitem{Jarlborg.83}
\bibinfo{author}{Jarlborg, T.}, \bibinfo{author}{Braun, H.} \&
  \bibinfo{author}{Peter, M.}
\newblock \bibinfo{title}{Structural properties and band structure of heavy
  fermion systems: {CeCu$_2$Si$_2$} and {LaCu$_2$Si$_2$}}.
\newblock \emph{\bibinfo{journal}{Z.~Phys.~B}} \textbf{\bibinfo{volume}{52}},
  \bibinfo{pages}{295--301} (\bibinfo{year}{1983}).

\bibitem{Roth.66}
\bibinfo{author}{Roth, L.}, \bibinfo{author}{Zeiger, H.} \&
  \bibinfo{author}{Kaplan, T.}
\newblock \bibinfo{title}{Generalization of the {Ruderman-Kittel-Kasuya-Yosida
  Interaction} for nonspherical {F}ermi surfaces}.
\newblock \emph{\bibinfo{journal}{Phys.~Rev.}} \textbf{\bibinfo{volume}{149}},
  \bibinfo{pages}{519--525} (\bibinfo{year}{1966}).

\bibitem{Stockert.04}
\bibinfo{author}{Stockert, O.} \emph{et~al.}
\newblock \bibinfo{title}{Nature of the {A} phase in {CeCu$_2$Si$_2$}}.
\newblock \emph{\bibinfo{journal}{Phys.~Rev.~Lett.}}
  \textbf{\bibinfo{volume}{92}}, \bibinfo{pages}{136401}
  (\bibinfo{year}{2004}).

\bibitem{Stock.08}
\bibinfo{author}{Stock, C.}, \bibinfo{author}{Broholm, C.},
  \bibinfo{author}{Hudis, J.}, \bibinfo{author}{Kang, H.~J.} \&
  \bibinfo{author}{Petrovic, C.}
\newblock \bibinfo{title}{Spin resonance in the d-wave superconductor
  {CeCoIn$_5$}}.
\newblock \emph{\bibinfo{journal}{Phys.~Rev.~Lett.}}
  \textbf{\bibinfo{volume}{100}}, \bibinfo{pages}{087001}
  (\bibinfo{year}{2008}).

\bibitem{Eremin.08}
\bibinfo{author}{Eremin, I.}, \bibinfo{author}{Zwicknagl, G.},
  \bibinfo{author}{Thalmeier, P.} \& \bibinfo{author}{Fulde, P.}
\newblock \bibinfo{title}{Feedback spin resonance in superconducting
  {CeCu$_2$Si$_2$ and CeCoIn$_5$}}.
\newblock \emph{\bibinfo{journal}{Phys.~Rev.~Lett.}}
  \textbf{\bibinfo{volume}{101}}, \bibinfo{pages}{187001}
  (\bibinfo{year}{2008}).

\bibitem{KirchnerandSi}
\bibinfo{author}{Kirchner, S.} \& \bibinfo{author}{Si, Q.}
\newblock \bibinfo{note}{To be published}.

\end{thebibliography}

\end{document}